\documentclass[12pt,a4paper]{article}
\usepackage{graphicx}
\usepackage[T1]{fontenc}
\usepackage[utf8]{inputenc}
\usepackage{textcomp}
\usepackage[sc,osf]{mathpazo}
\usepackage{a4wide}  
\usepackage{latexsym,amsthm,amsfonts,amsmath,mathrsfs,amssymb}
\usepackage{dsfont}
\usepackage{accents}
\usepackage[nosort]{cite}
\usepackage{booktabs} 
\usepackage[unicode,implicit]{hyperref}
\hypersetup{%
  pdftitle    = {Non-extremal, $\alpha'$-corrected black holes in 5-dimensional
    Heterotic Superstring Theory}
  pdfkeywords = {black hole, string theory, entropy, temperature,
    thermodynamics, Wald, Hawking, Smarr,  branes, supergravity, supersymmetry, higher order, corrections,  heterotic superstring}
  pdfauthor   = {Pablo A. Cano, Tom\'as Ort\'{\i}n, Alejandro Ruip\'erez and
    Matteo Zatti},
  plainpages  = true,
  colorlinks  = true,
  citecolor   = blue,
  urlcolor    = red,
  linkcolor   = black
}
\newcommand{\hepth}[1]{{\tt
\href{http://www.arXiv.org/abs/hep-th/#1}{hep-th/#1}}}
\newcommand{\grqc}[1]{{\tt
\href{http://www.arXiv.org/abs/gr-qc/#1}{gr-qc/#1}}}

\newcommand{\arxiv}[1]{{\tt arXiv:\href{http://www.arXiv.org/abs/#1}{#1}}}
\newcommand{\doi}[1]{{\tt DOI:\href{http://dx.doi.org/#1}{#1}}}

\allowdisplaybreaks

\makeatletter
\@addtoreset{equation}{section}
\makeatother

\begin{document}

\begin{flushright}
\small
IFT-UAM/CSIC-22-122\\
October 5\textsuperscript{th}, 2022\\
\normalsize
\end{flushright}

\vspace{1cm}

\begin{center}

  {\Large {\bf Non-extremal, $\alpha'$-corrected black holes\\[.5cm]
      in 5-dimensional Heterotic Superstring Theory}}

\vspace{1cm}

\renewcommand{\thefootnote}{\alph{footnote}}

{\sl\large Pablo A.~Cano,}$^{1,}$\footnote{Email: {\tt pabloantonio.cano[at]kuleuven.be}}
{\sl\large  Tom\'{a}s Ort\'{\i}n,}$^{2,}$\footnote{Email: {\tt tomas.ortin[at]csic.es}}
{\sl\large Alejandro Ruip\'erez,}$^{3,4,}$\footnote{Email: {\tt alejandro.ruiperez[at]pd.infn.it}}
{\sl\large and Matteo Zatti}$^{2,}$\footnote{Email: {\tt matteo.zatti[at]estudiante.uam.es}}

\setcounter{footnote}{0}
\renewcommand{\thefootnote}{\arabic{footnote}}

\vspace{0.8cm}

${}^{1}${\it Instituut voor Theoretische Fysica, KU Leuven\\
	Celestijnenlaan 200D, B-3001 Leuven, Belgium}\\

\vspace{0.4cm}

${}^{2}${\it Instituto de F\'{\i}sica Te\'orica UAM/CSIC\\
C/ Nicol\'as Cabrera, 13--15,  C.U.~Cantoblanco, E-28049 Madrid, Spain}

\vspace{0.4cm}

${}^{3}${\it Dipartimento di Fisica ed Astronomia ``Galileo Galilei'',\\
  Universit\`a di Padova, Via Marzolo 8, 35131 Padova, Italy}

\vspace{0.4cm}

${}^{4}${\it INFN, Sezione di Padova,Via Marzolo 8, 35131 Padova, Italy}

\vspace{0.7cm}


{\bf Abstract}
\end{center}
\begin{quotation}
  {\small We compute the first-order $\alpha'$ corrections of the non-extremal
    Strominger-Vafa black hole and its non-supersymmetric counterparts in the
    framework of the Bergshoeff-de Roo formulation of the heterotic
    superstring effective action. The solution passes several tests: its
    extremal limit is the one found in an earlier publication and the effect of
    a T~duality transformation on it is another solution of the same form with
    T~dual charges. We compute the Hawking temperature and Wald entropy
    showing that they are related by the first law and Smarr formula. On the
    other hand, these two contain additional terms in which the dimensionful
    parameter $\alpha'$ plays the role of thermodynamical variable.}
\end{quotation}

\newpage
\pagestyle{plain}

\tableofcontents

\newpage

\section{Introduction}

The microscopic interpretation of the entropy of a 5-dimensional extremal
black-hole solution of the $0^{\rm th}$-order in $\alpha'$ string effective
action by Strominger and Vafa in Ref.~\cite{Strominger:1996sh} is widely
acknowledged as one of the main successes of Superstring Theory. Since then,
extending this result beyond leading order in $\alpha'$ was a
must.\footnote{The problem of finding $\alpha'$ corrections to the classical
  black-hole solutions had been considered before in
  Refs.~\cite{Campbell:1990ai,Campbell:1991kz,Natsuume:1994hd}.}  From the
macroscopic side, advances in this front were possible thanks to the
construction in \cite{Hanaki:2006pj} of a four-derivative supersymmetric
invariant containing the mixed gauge/gravitational Chern-Simons term,
$A\wedge {\rm{tr}}\left(R\wedge R\right)$, and to the development of the
entropy function formalism by Sen \cite{Sen:2007qy}, which allows to extract
the microcanonical form of the entropy from corrections to the near-horizon
geometry, which are much easier to study. Making profit of this,
Refs.~\cite{Castro:2007hc, Castro:2007ci, Castro:2008ne} calculated
corrections to the macroscopic entropy of asymptotically-flat 5-dimensional
black holes. These were soon later matched from a microscopic analysis
\cite{Castro:2008ys} for the particular supergravity model arising from type
IIB on K$_{3}\times {\mathrm S}^1$ (or, equivalently, heterotic string theory
on ${\mathrm T}^{5}$), extending the validity of the pioneering result of
\cite{Strominger:1996sh} beyond the supergravity regime. Subsequent
calculations in the dual heterotic frame were later performed in
\cite{Prester:2008iu}, being also in agreement with the microscopic entropy
previously derived in \cite{Castro:2008ys}.

In spite of these successes, there were some open issues at that time that
have not been fully addressed until very recently. One of them was to actually
show that there exists a corrected solution extending from the
AdS${}_{2}\times S^3$ near-horizon geometry to the asymptotically-flat
region. Such solution was obtained in \cite{Cano:2018qev} in a fully
analytical form. Having this solution at our disposal allows us to study
aspects of the solutions that are not fully accessible within the near-horizon
description. An instance of this is the proper identification between the
conserved charges of the solution and the parameters that describe the
microscopic system. As discussed in \cite{Cano:2018hut, Cano:2021dyy}, this
can lead to important consequences in some specific cases.

Another important issue that was not put on a solid ground is the macroscopic
calculation of the entropy. This is due to the fact that the presence of
Chern-Simons terms does not allow for a direct application of Wald's entropy
formula \cite{ Lee:1990nz,Wald:1993nt,Iyer:1994ys}, as it is well known
\cite{Tachikawa:2006sz, Elgood:2020svt, Elgood:2020mdx,
  Elgood:2020xwu,Elgood:2020nls}.\footnote{It is not difficult to see that,
  due to the Lorentz-Chern-Simons terms present in the Kalb-Ramond field
  strength, the Iyer-Wald prescription \cite{Iyer:1994ys} gives a
  frame-dependent entropy formula. The coefficients are not correct, either.}
Strictly speaking, this issue is not specific of Wald's formula, as it also
shows up in the entropy function formalism \cite{Sen:2007qy,
  Sahoo:2006pm}. Lacking a generalization of the Iyer-Wald prescription, an
strategy to deal with the Chern-Simons terms present in the heterotic
effective action was proposed in \cite{Sahoo:2006pm}, being successfully
applied to compute the entropy of 5-dimensional 3-charge black holes in
\cite{Prester:2008iu, Faedo:2019xii}. Remarkably, a gauge-invariant formula
for the heterotic superstring effective action has been very recently derived
in Ref.~\cite{Elgood:2020nls}. Making use of it, the macroscopic entropy of
the 5-dimensional extremal Strominger-Vafa and its non-supersymmetric
analogues was finally computed in Ref.~\cite{Cano:2021nzo}, further confirming
the existing results and putting them on a much more solid ground.

The entropy formula obtained in Ref.~\cite{Elgood:2020nls} is manifestly frame
independent. In certain frames it reduces to a formula similar to the one
obtained applying the Iyer-Wald prescription, but with different numerical
coefficients which lead to a different result. When applied to the
$\alpha'$-corrected non-extremal Reissner-Nordstr\"om black hole in
Ref.~\cite{Cano:2019ycn} only the formula obtained in
Ref.~\cite{Elgood:2020nls} satisfies the thermodynamical relation
$\partial S/\partial M = 1/T$ which follows from the first law of black-hole
thermodynamics. It is worth stressing that the main justification for the
identification of the Noether charge associated to the Killing vector that
generates the event horizon with the black-hole entropy is the fact that it
satisfies the first law \cite{Wald:1993nt,Iyer:1994ys} and that it reduces to
the Bekenstein-Hawking entropy in absence of higher-order terms. A general
proof of the second law of black-hole thermodynamics for the Wald entropy is
not yet available, although there has been remarkable progress in this
direction \cite{Wall:2015raa,Hollands:2022fkn}.

Although the test of the entropy formula of Ref.~\cite{Elgood:2020nls}
performed in Ref.~\cite{Cano:2019ycn} is quite convincing, it would be most
desirable to perform a similar test using a non-extremal generalization of the
Strominger-Vafa black hole checking also that in the extremal limit one
recovers the entropy computed in Ref.~\cite{Cano:2021nzo} directly form the
extremal solution. The non-extremal generalization of the Strominger-Vafa
black hole was found in Ref.~ \cite{Horowitz:1996ay} shortly after the
extremal one but its $\alpha'$ corrections have never been
computed.\footnote{Typically, the near-horizon limit non-extremal black holes
  is not a solution and this approximation cannot be made in this case.} The
goal of this work is to compute the $1^{\rm st}$-order corrections to the
non-extremal Strominger-Vafa black hole and its main thermodynamical
properties. To this end, as in our previous works
Refs.~\cite{Cano:2018qev,Chimento:2018kop,Cano:2018brq,Cano:2019oma,Cano:2021nzo,Ortin:2021win},
we are going to use the Bergshoeff-de Roo version of the heterotic superstring
effective action \cite{Bergshoeff:1989de} because its supersymmetrization is
known among other reasons.\footnote{Earlier results on this effective action
can be found in Refs.~\cite{Gross:1986mw,Metsaev:1987zx} and more recent
  work in Refs.~\cite{Chemissany:2007he,Baron:2017dvb,Baron:2018lve}.}

Computing $\alpha'$ corrections is a complicated problem: as usual, a good
ansatz compatible with the symmetry of the problem is needed, but the number
of independent functions that one still has to determine can be quite
large.\footnote{Smaller numbers lead to equations which are incompatible or
  which are very difficult to solve, as one finds by trial and error.} As we
pointed out in Ref.~\cite{Cano:2021nzo}, T~duality can be used to constrain
the form of the corrections at least in the solution at hands, which is
expected to be related to itself (up to redefinitions of the integration
constants) under some T~duality transformations.

T~duality, which arises in string theory in the presence of toroidal compact
directions, takes its simplest form when it is expressed in terms of the
fields that result from the dimensional reduction of the theory over those
compact directions. In this way, as shown in Ref.~\cite{Bergshoeff:1994dg} it
is easy to recover the well-known Buscher rules
\cite{Buscher:1987sk,Buscher:1987qj}. This observation has subsequently been
used to extend the Buscher rules to the type~II theories
\cite{Bergshoeff:1995as,Meessen:1998qm} and to the heterotic superstring
effective action to $1^{\rm st}$ order in $\alpha'$
\cite{Bergshoeff:1995cg,Elgood:2020xwu}. The possibility to constrain some of
the corrections follows from this observation: all the fields that descend
from the Kalb-Ramond field via dimensional reduction receive explicit
$\alpha'$ corrections that can be computed at $1^{\rm st}$-order using the
$0^{\rm th}$-order solution \cite{Elgood:2020xwu}. These explicit corrections
are mixed and interchanged with implicit corrections by T~duality and,
demanding invariance, one finds relations between the known (explicit)
corrections and the unknown (implicit) ones that can be used to determine the
latter. In our case, these relations have prove crucial to find all the
corrections.

This paper is organized as follows: in Section~\ref{sec-10dimensional} we
describe our 10-dimensional ansatz and in Section~\ref{sec-solving} we explain
how we have solved them using the T~duality constraints mentioned above. In
Section~\ref{sec-thermodynamics} we compute the basic thermodynamical
quantities of the corrected solution and in Section~\ref{sec-conclusions} we
present our conclusions and point to some directions for future work. The
Appendices contain complementary information: the $1^{\rm st}$ order in
$\alpha'$ heterotic superstring effective action and the equations of motion
of the fields to that order are reviewed in Appendix~\ref{app-action} in the
conventions we are using. The $0^{\rm th}$-order solutions we start from are
reviewed in Appendix~\ref{app-zerothorder}. In particular, we derive its basic
thermodynamical properties which have to be recovered in the
$\alpha'\rightarrow 0$ limit. The first law of black hole mechanics is checked
at $0^{\rm th}$ order in Appendix~\ref{sec-checkingfirstlawzeroth}.

\section{The 10-dimensional ansatz}
\label{sec-10dimensional}

Based on our knowledge of the $0^{\rm th}$-order non-extremal
solution\footnote{This solutions, found in Ref.~\cite{Horowitz:1996ay}, is
  reviewed in Appendix~\ref{app-zerothorder}. Some of the notation used here
  is introduced there. The reader is encouraged to read that appendix first.}
and on our knowledge of the first-order extremal solutions
\cite{Cano:2018qev,Cano:2018brq,Cano:2019oma,Cano:2021nzo}, an educated ansatz
for the 3-charge 5-dimensional black-hole solution that we want to find and
study turns out to require the introduction of 7 independent functions of
$\rho$

\begin{equation}
  \label{eq:7functions}
  \mathcal{Z}_{0}\,,\,\,\mathcal{Z}_{+}\,,\,\,\mathcal{Z}_{-}\,,\,\,
  \mathcal{Z}_{h0}\,,\,\,\mathcal{Z}_{h-}\,,\,\,W_{tt}\,,\,\,W_{\rho\rho}\,.
\end{equation}

\noindent
The new functions $\mathcal{Z}_{h0}$ and $\mathcal{Z}_{h-}$ are, respectively,
identical to $\mathcal{Z}_{0}$ and $\mathcal{Z}_{-}$ at $0^{\rm th}$ order and they
have to be introduced because these functions get different corrections when
they occur in different components of the fields of the solutions. The
functions $W_{tt}$ and $W_{\rho\rho}$ reduce to $W$ at $0^{\rm th}$ order and are
needed because $W$ gets different corrections when it is part of the $tt$ or
the $\rho\rho$ components of the metric.

As for the constants, including the integration constants found at $0^{\rm th}$
order, we now need 10:

\begin{equation}
  c_{\hat{\phi}}\,,\,\,\beta_{0}\,,\,\,\beta_{+}\,,\,\,\beta_{-}\,,\,\,
  q_{0}\,,\,\,q_{+}\,,\,\,q_{-}\,,\,\,\omega\,,\,\,
  \hat{\phi}_{\infty}\,,\,\,k_{\infty}\,.
\end{equation}

We expect the same number of independent physical parameters as in the
$0^{\rm th}$-order solutions, namely 6, which means that we need 4 relations between
these constants. We are going to assume that the three $0^{\rm th}$-order relations
Eqs.~(\ref{eq:omegaqbetarelation}) are satisfied at first order as well,
without corrections.  A fourth relation involving $c_{\hat{\phi}}$ is obtained
when one imposes that the asymptotic value of the dilaton is
$ \hat{\phi}_{\infty}$.

The functions in Eq.~(\ref{eq:7functions}) are assumed to have the following
form ($i=0,+,-$, $j=tt, \rho\rho$):

\begin{equation}
  \begin{aligned}
    \mathcal{Z}_{i}
    & =
    1 + \frac{q_{i}}{\rho^{2}} + \alpha' \delta \mathcal{Z}_{i}\,,
    \\
    & \\
    \mathcal{Z}_{hi}
    & =
    1 + \frac{q_{i}}{\rho^{2}} + \alpha' \delta \mathcal{Z}_{hi}\,,
    \\
    & \\
    W_{j}
    & =
    1+\frac{\omega}{\rho^{2}} + \alpha' \delta W_{j}\,.
  \end{aligned}
\end{equation}

\noindent
Thus, they become the functions of the $0^{\rm th}$-order ansatz
Eqs.~(\ref{eq:Zs3-chargezerothorder}) when $\alpha'=0$
(note that $\mathcal{Z}_{h0}=\mathcal{Z}_{0}$, $\mathcal{Z}_{h-}=\mathcal{Z}_{-}$ and $W_{tt}=W_{\rho\rho}=W$).

In terms of these functions and constants, the 10-dimensional fields are
assumed to be given by

\begin{subequations}
\label{eq:3-charge10dsolution}
\begin{align}
d\hat{s}^{2}
  & =
    \frac{1}{\mathcal{Z}_{+}\mathcal{Z}_{-}}W_{tt}dt^{2}
    -\mathcal{Z}_{0}(W_{\rho\rho}^{-1}d\rho^{2}+\rho^{2}d\Omega_{(3)}^{2})
    \nonumber \\
  & \nonumber \\
  & 
    -\frac{k_{\infty}^{2}\mathcal{Z}_{+}}{\mathcal{Z}_{-}}
    \left[dz+\beta_{+}k_{\infty}^{-1}
    \left(\mathcal{Z}^{-1}_{+}-1\right)dt\right]^{2}
    -dy^{m}dy^{m}\,,
\hspace{.5cm}
m=1,\ldots,4\,,
\label{eq:d10metric}
  \\
  & \nonumber \\
  \hat{H}^{(1)}
  & = 
    \beta_{-}d\left[ k_{\infty}\left(\mathcal{Z}^{-1}_{h-}-1\right)
    dt \wedge dz\right]
    +\beta_{0}\rho^{3}\mathcal{Z}'_{h0}\omega_{(3)}\,,
  \\
  & \nonumber \\
  e^{-2\hat{\phi}}
  & =
    -\frac{2 c_{\hat{\phi}}}{\rho^{3}}\frac{\mathcal{Z}_{h-}^{2}}{ \mathcal{Z}_{0} \mathcal{Z}_{-}\mathcal{Z}_{h-}'}   \sqrt{W_{tt}/W_{\rho\rho}}\,.
\end{align}
\end{subequations}

This ansatz reduces to the $0^{\rm th}$-order one
Eqs.~(\ref{eq:3-charge10dsolutionzerothorder}) when
$\mathcal{Z}_{h0}=\mathcal{Z}_{0}$, $\mathcal{Z}_{h-}=\mathcal{Z}_{-}$ and
$W_{tt}=W_{\rho\rho}=W$.

All we have to do now is to substitute this ansatz into the equations of
motion derived in Appendix~\ref{sec-eom} and solve for the corrections
$\delta \mathcal{Z}_{i},\delta \mathcal{Z}_{hi},\delta W_{j}$. Needless to
say, this is a very difficult task.  In the next section, we explain step by
step how we have done it.

\section{Solving the equations of motion}
\label{sec-solving}

We start by the equations which are the simplest to solve: the Kalb-Ramond
(KR) equation (\ref{eq:Krequation}) and Bianchi identity
Eq.~(\ref{eq:bianchi}).\footnote{Our ansatz is written in terms of the KR
  field strength and, therefore, we must impose the KR Bianchi identity. } As
a matter of fact, the ansatz for the dilaton has been chosen in such a way
that the KR equation (\ref{eq:Krequation}) is automatically solved.

On the other hand, the only non-trivial component of the Bianchi identity
Eq.~(\ref{eq:bianchi}) to first order in $\alpha'$ is the $r\theta\phi\psi$
and, using the $0^{\rm th}$-order solution, it becomes a differential equation for
$\delta\mathcal{Z}_{h0}$ which is solved by

\begin{equation}
  \label{eq:deltaZh0}
  \delta\mathcal{Z}_{h0}
  =   d^{(0)}_{h0} +\frac{d^{(2)}_{h0}}{\rho^{2}}
  +\frac{2 q_{0}^{3} + \omega \left(q_{0}^{2} + 9 q_{0} \rho^{2} + 6
      \rho^{4}\right)}{2 q_{0} \rho^{2} (q_{0} + \rho^{2})^{2}}
  -\frac{3 \omega}{q_{0}^{2}}\log\left(1+\frac{q_{0}}{\rho^{2}}\right)\,,
\end{equation}

\noindent
where $d^{(0)}_{h0}$ and $d^{(2)}_{h0}$ are integration constants. Imposing
that $\mathcal{Z}_{h0}$ does not receive $\alpha'$ corrections at infinity we
obtain $d^{(0)}_{h0} = 0$.

The remaining equations of motion for the 6 remaining functions form a highly coupled system of 5
independent differential equations. Note that we have more unknown functions than equations, and the reason is that our ansatz contains some gauge freedom: one of the functions can be chosen at will by means of an infinitesimal transformation of the radial coordinate. This freedom will prove to be very useful to express the solution in a simple form.
Despite the complexity of the equations, they can solved by the
following procedure: first, we use as an ansatz for the $\delta \mathcal{Z}$s
and $\delta W$s the following series with arbitrary coefficients

\begin{equation}
  \label{eq:seriesdef}
  \delta \mathcal{Z}_{i} = \sum_{n = 1} \frac{d^{(2n)}_{i}}{\rho^{2n}}\,,
  \quad
  \delta \mathcal{Z}_{hi} = \sum_{n = 1} \frac{d^{(2n)}_{hi}}{\rho^{2n}}\,,
  \quad
  \delta W_{j} = \sum_{n = 1} \frac{d^{(2n)}_{wj}}{\rho^{2n}}\,.
\end{equation} 

Notice that we have assumed that all the powers in $1/\rho$ are even and that
there is no correction to the asymptotic values of $\mathcal{Z}$s and
$W$s. Furthermore, it follows that the coefficient $c_{\hat{\phi}}$ that sets $\lim_{r\rightarrow\infty} \hat\phi =\hat\phi_{\infty}$ is given by 

\begin{equation}
  c_{\hat{\phi}}
  =
  \left(q_{-} + \alpha' d^{(2)}_{h-}\right) e^{-2\hat{\phi}_{\infty}}\,.
\end{equation}

Plugging this ansatz into the Einstein equations (\ref{eq:Einsteinequation})
and the dilaton equation (\ref{eq:dilatonequation}) and demanding that the
they are solved order by order in powers of of $1/\rho$ one obtains algebraic
equations for the coefficients $d^{(2n)}$s that one can solve for a large
value of $n$.  It turns out that not all of these coefficients are determined
by these equations: the equations of motion are solved for arbitrary values of
the coefficients $d^{(2)}_{i}$, $d^{(2)}_{hi}$, $d_{wj}^{(2)}$ and
$d^{(2n)}_{h-}$ with $n\ge 3$. Having determined the coefficients for a large
enough value of $n$ we can determine the functions associated to those
expansions resuming the series\footnote{Actually ,we have used a symbolic
  manipulation program that makes a good guess for those functions based on a
  very large number of terms of the series.}. Finally, we have to check that
those functions do solve exactly the equations of motion to first order in
$\alpha'$.

Before carrying out this program, though, it is convenient to make some
considerations concerning T~duality.

As discussed in Appendix~\ref{sec-Tdualityzeroth}, the $0^{\rm th}$-order solution,
understood as a family, is invariant under the $0^{\rm th}$-order Buscher T~duality
transformations \cite{Buscher:1987sk,Buscher:1987qj}: one member of the family
is transformed into another with different values of the parameters. It is
enough to transform the parameters according to the rules in
Eq.~(\ref{eq:zerothorderTdualityparameters}). At $0^{\rm th}$ order in $\alpha'$,
the physical interpretation of the transformations of the parameters is the
expected one: momentum and winding are interchanged and the radius of the
compact dimension is inverted.

It is reasonable to expect that the T~duality invariance of the family of
solutions is preserved at first order in $\alpha'$ when the
$\alpha'$-corrected Buscher T~duality transformations derived in
Refs.~\cite{Bergshoeff:1995cg,Elgood:2020xwu} are used because we are just
dealing with a more accurate description of exactly the same physical
system. We have shown in Ref.~\cite{Cano:2021nzo} that, in the extremal limit,
this is what happens and there is no reason to expect otherwise in the
non-extremal case.

T~duality is best studied in the dimensionally-reduced solution, where it
takes the form of a discrete symmetry transformation of the action
\cite{Bergshoeff:1994dg} that, essentially, interchanges two vector fields and
inverts a scalar. At $0^{\rm th}$ order, these transformations are given in
Eq.~(\ref{eq:Tdualityzerothorder}).

At first order in $\alpha'$ the transformations have the same
form:\footnote{We use $C^{(1)}$ for the winding vector to distinguish it from
  the 2-form $B^{(1)}$. Its 2-form field strength is $G^{(1)}=dC^{(1)}$.}

\begin{equation}
  \label{eq:Tdualityfirstorder}
  A \leftrightarrow C^{(1)}\,,
  \hspace{1cm}
  k \leftrightarrow 1/k^{(1)}\,.
\end{equation}

The main difference with the $0^{\rm th}$-order ones is that the relation between the
winding vector $C^{(1)}$ and the higher-dimensional fields is modified by
$\alpha'$ corrections (hence the upper $(1)$ index). The scalar $k^{(1)}$
contains $\alpha'$ corrections as well. Then, the first-order solution can
only be self-dual if the $\alpha'$ corrections one finds for $A$ and $k$ are
related to the explicit $\alpha'$ corrections of $C^{(1)}$ and $k^{(1)}$ in a
very specific way Ref.~\cite{Cano:2021nzo}. Thus, the expected T~duality
invariance of the $\alpha'$-corrected solution can be used to simplify the
problem of finding the corrections and also to test them.

In order to exploit this expected T~duality invariance of the corrected
solutions it is convenient to start by finding their 5-dimensional form using
the relations between higher- and lower-dimensional fields found in
Ref.~\cite{Elgood:2020xwu}.

\subsection{The 5-dimensional form}
\label{sec-d5-5dimensionalcorrected}

In order to find the 5-dimensional fields we first need the 10-dimensional KR
2-form potential. Its existence is guaranteed by the fact that we have solved
the Bianchi identity determining $\delta\mathcal{Z}_{h0}$ in
Eq.~(\ref{eq:deltaZh0}).  The KR 2-form is given by

\begin{equation}
  \hat{B}^{(1)}
  =
  \beta_{-}k_{\infty}\left(\mathcal{Z}^{-1}_{h-}-1\right)dt\wedge dz
  +\tfrac{1}{4}\left[c+\alpha'd+ (\beta_{0}q_{0}+\alpha')\cos{\theta}\right]d\phi\wedge
  d\psi\,,
\end{equation}

\noindent
where $c$ and $d$ are integration constants whose values can be chosen in
different coordinate patches so as to make the field globally well-defined.
Locally, they can be set to zero, and we will do so from now on for the sake
of simplicity.

With this result we can compute the 5-dimensional KR 2-form $B^{(1)}$ and
winding vector $C^{(1)}$ using the relations \cite{Elgood:2020xwu}

\begin{subequations}
  \begin{align}
    B^{(1)}{}_{\mu\nu}
    & =
    \hat{B}_{\mu\nu}
    +\hat{g}_{\underline{z}[\mu}
         \hat{B}_{|\nu]\, \underline{z}}
      +\frac{\alpha'}{4}\hat{g}_{\underline{z}[\mu}\hat{\Omega}^{(0)}_{(-)\, |\nu]}{}^{\hat{a}}{}_{\hat{b}}
      \hat{\Omega}^{(0)}_{(-)\, \underline{z}}{}^{\hat{b}}{}_{\hat{a}}
    /\hat{g}_{\underline{z}\underline{z}}\,,
    \\
    & \nonumber \\
    C^{(1)}{}_{\mu}
    & =
     \hat{B}_{\mu\, \underline{z}}
      -\frac{\alpha'}{4}\left[
      \hat{\Omega}^{(0)}_{(-)\, \mu}{}^{\hat{a}}{}_{\hat{b}}
      \hat{\Omega}^{(0)}_{(-)\, \underline{z}}{}^{\hat{b}}{}_{\hat{a}}
      -\hat{g}_{\mu \underline{z}}\hat{\Omega}^{(0)}_{(-)\, \underline{z}}{}^{\hat{a}}{}_{\hat{b}}
      \hat{\Omega}^{(0)}_{(-)\, \underline{z}}{}^{\hat{b}}{}_{\hat{a}}
  /\hat{g}_{\underline{z}\underline{z}}\right]\,,
  \end{align}
\end{subequations}

\noindent
and the scalar combination

\begin{equation}
  k^{(1)}
=
  k\left[1-\frac{\alpha'}{4}\hat{\Omega}^{(0)}_{(-)\,
      \underline{z}}{}^{\hat{a}}{}_{\hat{b}}\hat{\Omega}^{(0)}_{(-)\,
      \underline{z}}{}^{\hat{b}}{}_{\hat{a}}/\hat{g}_{\underline{z}\underline{z}}\right]\,.
\end{equation}

The result, including the 5-dimensional KR 2-form potential and the scalar
combination $k^{(1)}$, can be expressed in the string frame as follows:

\begin{subequations}
\label{eq:5dsolution1storder}
\begin{align}
  ds^{2}
  & =
\frac{W_{tt}}{\mathcal{Z}_{+}\mathcal{Z}_{-}}dt^{2}
    -\mathcal{Z}_{0}
    \left({W}^{-1}_{\rho\rho}d\rho^{2}+\rho^{2}d\Omega^{2}_{(3)}\right)\,,
  \\
  & \nonumber \\
  \label{eqM015}
  H^{(1)}
  & =
    \beta_{0} \rho^{3} \mathcal{Z}_{h0}'\,\omega_{(3)}\,, 
  \\
  & \nonumber \\
   \label{eqM010}
  k
  & =
    k_{\infty} \sqrt{\mathcal{Z}_{+}/\mathcal{Z}_{-}}\,,
  \\
  & \nonumber \\
   \label{eqM011}
  k^{(1)}
  & =
    k_{\infty} \sqrt{\mathcal{Z}_{+}/\mathcal{Z}_{-}}
    \left(1+\alpha' \Delta_{k}\right)\,,
  \\
  & \nonumber \\
   \label{eqM012}
  A
  & =
    k_{\infty}^{-1}\beta_{+}\left[-1+\frac{1}{\mathcal{Z}_{+}}\right]dt \,,
  \\
  & \nonumber \\
   \label{eqM013}
  C^{(1)}
  & =
    k_{\infty}\beta_{-}\left[-1+\frac{1}{\mathcal{Z}_{h-}}
    \left(1+\alpha'\frac{\Delta_{C}}{\beta_{-}}\right)\right]
    dt \,,
  \\
  & \nonumber \\
   \label{eqM005}
  e^{-2\phi}
  & =
    - \frac{2\,c_{\phi}}{\rho^{3} \mathcal{Z}_{h-}'}
    \frac{\mathcal{Z}_{h-}^{2}}{\mathcal{Z}_{-}^{2}}
    \sqrt{\frac{W_{tt}}{W_{\rho\rho}}}
    \frac{\sqrt{\mathcal{Z}_{+}\mathcal{Z}_{-}}}{\mathcal{Z}_{0}} \,,
\end{align}
\end{subequations}

\noindent
where

\begin{equation}
  c_{\phi}
  =
  c_{\hat{ \phi}}\, k_{\infty}\,,  
\end{equation}

\noindent
and where the functions $\Delta_{k}$ and $\Delta_{C}$ are given by

\begin{subequations}
  \begin{align}
    \label{eq:Deltak}
    \Delta_{k}
    & =
      \frac{-W\left(\mathcal{Z}_{+}\mathcal{Z}_{-}'-\mathcal{Z}_{-}\mathcal{Z}_{+}'\right)^{2}+\left(\beta_{-}\mathcal{Z}_{+}\mathcal{Z}_{-}'+\beta_{+}\mathcal{Z}_{-}\mathcal{Z}_{+}'\right)^{2}}{8\,\mathcal{Z}_{0}\mathcal{Z}_{-}^{2}\mathcal{Z}_{+}^{2}}\,,
    \\
    & \nonumber \\
    \Delta_{C}
    & =
      \frac{-\left(\beta_{-}\mathcal{Z}_{+}\mathcal{Z}_{-}'+\beta_{+}\mathcal{Z}_{+}'\mathcal{Z}_{-}\right)W'+2\left(\beta_{+}+\beta_{-}\right)\mathcal{Z}_{+}'W\mathcal{Z}_{-}'}{8\,\mathcal{Z}_{0}\mathcal{Z}_{-}\mathcal{Z}_{+}}\,.
\end{align}
\end{subequations}

These functions contain the explicit $\alpha'$ corrections to the fields
$C^{(1)}$ and $k^{(1)}$ we have mentioned before.

We still have to do two additional modifications.  First, the modified
Einstein-frame metric is given by

\begin{equation}
  \label{eq:modifiedEinsteinmetriccorrected}
  \begin{aligned}
    ds^{2}_{E} & = \left[ -\frac{2c_{\phi} e^{2\phi_{\infty}}
        \mathcal{Z}_{h-}^{2}}{\rho^{3}\mathcal{Z}_{h-}'\mathcal{Z}_{-}^{2}}
    \right]^{2/3} \left(\frac{W_{tt}}{W_{\rho\rho}}\right)^{1/3}
    \left\{f^{2}W_{tt}dt^{2}
      -f^{-1}\left(W_{\rho\rho}^{-1}d\rho^{2}+\rho^{2}d\Omega_{(3)}^{2}\right)\right\}\,,
    \\
    & \\
    f^{-3}
    & =
    \mathcal{Z}_{+}\mathcal{Z}_{-}\mathcal{Z}_{0}\,.
  \end{aligned}
\end{equation}

Second, since the equation of motion of the KR field takes exactly the same
form as in the $0^{\rm th}$-order case we can use the same duality relation
Eq.~(\ref{eq:HdualK}) to convert the 3-form field strength into the 2-form
field strength

\begin{equation}
  K
  =
    d\left[-\beta_{0}\left(\mathcal{Z}^{-1}_{h0}-1\right)dt\right]\,.
\end{equation}

The action of a T~duality transformation in the direction of the compact
coordinate $z$ on the solution is just given by
Eq.~(\ref{eq:Tdualityfirstorder}) with the rest of the fields being invariant.
As we have discussed, the family of solutions that we are after is expected to
be invariant under these transformations.

We are now ready to apply the procedure we have outlined above.

\subsection{A particular solution}
\label{sec-particularsolution}

First, for the sake of convenience, we want to re-sum the series in the
simplest case, setting $d^{(2)}_{i} = d^{(2)}_{hi} = d_{wj}^{(2)} = 0$ and
$d^{(2n)}_{h-} = 0$ for $n\ge 3$. We obtain the following values for the
corrections:\footnote{In these expressions we have used the functions
  $\mathcal{Z}_{i}$, $\mathcal{Z}_{hi}$, $W_{j}$ to get simpler
  expressions. Since the corrections $\delta \mathcal{Z}_{i}$,
  $\delta \mathcal{Z}_{hi}$, $\delta W_{j}$ are multiplied by $\alpha'$, only
  the zeroth-order term of these functions actually contributes to these
  expressions and at $0^{\rm th}$ order
  $\mathcal{Z}_{hi}=\mathcal{Z}_{i},W_{j}=W$. }

\begin{subequations}
  \label{eq001}
\begin{align}
          \delta \mathcal{Z}_{h0}
          & =
          \frac{2 q_{0}^{3} + \omega \left(q_{0}^{2} + 9 q_{0} \rho^{2} + 6
            \rho^{4}\right)}{2 q_{0} \rho^{2} (q_{0} + \rho^{2})^{2}} - \frac{3
            \omega}{q_{0}^{2}}\log{\mathcal{Z}_{0}}\,,
  \\
  & \nonumber \\
	\begin{split}
          \delta \mathcal{Z}_{0} & = \frac{-3 q_{0}^{4} \left(2
              \rho^{2}+\omega\right)+5 q_{0}^{3} \omega \left(3
              \rho^{2}+\omega\right)+\rho^{4} \omega^{2} \left(\rho^{2}+2
              \omega\right)}{4 q_{0} \rho^{2} \left(q_{0}+\rho^{2}\right)^{2}
            (q_{0}-\omega)^{2}}
          \\
          &  \\
          & \hspace{.5cm} \frac{+q_{0} \rho^{2} \omega \left(-3 \rho^{4}-5 \rho^{2}
              \omega+4 \omega^{2}\right) +q_{0}^{2} \left(2 \rho^{6}+3 \rho^{4}
              \omega-13 \rho^{2} \omega^{2}-2 \omega^{3}\right)}{4 q_{0} \rho^{2}
            \left(q_{0}+\rho^{2}\right)^{2} (q_{0}-\omega)^{2}}
          \\
          & \\
          & \hspace{.5cm} +\frac{\omega^{2} q_{0} \left(2
              \rho^{2}+\omega\right)-\rho^{2} \omega^{2}
            \left(\rho^{2}+2 \omega\right) }{4 q_{0}^{2} \rho^{2}
            (q_{0}-\omega)^{2}}\log{\mathcal{Z}_{0}}
          \\
          & \\
          & \hspace{.5cm} +\frac{2 q_{0}^{2} \left(2 \rho^{2}+\omega\right)+3
            \rho^{2} \omega \left(\rho^{2}+2 \omega\right)-q_{0} \left(2 \rho^{4}+10
              \rho^{2} \omega+3 \omega^{2}\right) }{4 \rho^{2} (q_{0}-\omega)^{2}
            \omega}\log{W}\,,
	\end{split}
  \\
  & \nonumber \\
  \delta\mathcal{Z}_{h-}
          & =
            0\,,
  \\
  & \nonumber \\
  \delta\mathcal{Z}_{-}
          & =
            \mathcal{Z}_{-}
            \left( \Delta_{k} - \frac{\Delta_{C}}{\beta_{+}}\right)\,,
  \\
  & \nonumber \\
  \delta\mathcal{Z}_{+}
          & =
            -\mathcal{Z}_{+}\frac{\Delta_{C}}{\beta_{+}} \,,
  \\
  & \nonumber \\
  \delta W_{tt}
          & =
            -\frac{\beta_{-}(W+\beta_{+}\beta_{-})+ W(\beta_{+} + \beta_{-})}{8 \beta_{+} \mathcal{Z}_{0} \mathcal{Z}_{-}} \mathcal{Z}_{-}' W' \,,
  \\
          & \nonumber \\
  \begin{split}
  \delta W_{rr}
          & =
            \frac{-\rho^{2} \omega^{2} \left(2 \rho^{4}+3 \rho^{2} \omega+\omega^{2}\right)+q_{0}^{3} \left(-4 \rho^{4}-4 \rho^{2} \omega+2 \omega^{2}\right)}{4 q_{0} \rho^{4}
              \left(q_{0}+\rho^{2}\right) (q_{0}-\omega)^{2}}
            \\
            & \\
            & \hspace{.5cm} +\frac{q_{0}^{2} \left(-4 \rho^{6}+2
		\rho^{4} \omega+6 \rho^{2} \omega^{2}-4 \omega^{3}\right)+q_{0} \omega \left(6 \rho^{6}+5 \rho^{4} \omega-\rho^{2} \omega^{2}+2 \omega^{3}\right)}{4 q_{0} \rho^{4}
              \left(q_{0}+\rho^{2}\right) (q_{0}-\omega)^{2}}
            \\
            & \\
            & \hspace{.5cm}
            +\frac{\omega^{2} \left(2 \rho^{4}+3 \rho^{2}
                \omega+\omega^{2}\right)}{4 q_{0}^{2} \rho^{2}
              (q_{0}-\omega)^{2}}  \log{\mathcal{Z}_{0}}
             +\frac{(2 q_{0}-3 \omega) \left(2 \rho^{4}+3 \rho^{2}
		\omega+\omega^{2}\right)}{4 \rho^{2} (q_{0}-\omega)^{2} \omega}\log{W}\,.
	\end{split} 
\end{align}
\end{subequations} 

This solution, however, turns out to have a singular horizon. This is due to
the choice of integration constants we have made,
($d^{(2)}_{i} = d^{(2)}_{hi} = d_{wj}^{(2)} = 0$ and $d^{(2n)}_{h-} = 0$ for
$n\ge 3$). In the next section we are going to fix this problem using this
particular solution and modifying it by allowing for other choices of the
integration constants.

\subsection{The general solution}

We now allow the coefficients of the $1/\rho^{2}$ terms in the expansions
Eq.~(\ref{eq:seriesdef}) to be different from zero and demand that the
solution has a regular black hole solution. We are going to consider
separately the $\omega>0$ and $\omega <0$ cases. At $0^{\rm th}$ order, these
two cases are equivalent up to a coordinate transformation but it may happen
that this equivalence is broken by the $\alpha'$ corrections. In that case we
would get two inequivalent solutions. While we do not really expect that
black-hole uniqueness/no-hair is broken by the $\alpha'$ corrections, there is
no actual theorem forbidding it and it must be explicitly checked.

\subsubsection{Horizon regularity for $\omega > 0$}

At $0^{\rm th}$ order the horizon is at $\rho = 0$ when $\omega > 0$. At first order
it may be shifted to $\rho_{H}=\alpha' \delta \rho$. Therefore, we perform an
expansion in $z = (\rho-\rho_{H})$, obtaining

\begin{subequations}
\begin{align}
  e^{-2\phi}
  & =
    y_{\phi}^{(0)} + \alpha' y_{\phi}^{(0,\log)} \log{z}
    +\mathcal{O}(z) \,,
  \\
  & \nonumber \\
  k
  & =
    y_{k}^{(0)} +  \alpha' y_{k}^{(0,\log)} \log{z} + \mathcal{O}(z) \,,
  \\
  & \nonumber \\
  g_{E\,tt}
  & =
    \alpha'y_{tt}^{(0)}+  \alpha'y_{tt}^{(1)}z+ y_{tt}^{(2)}z^{2} +
    \alpha'y_{tt}^{(2,\log)}z^{2}\log{z} + \mathcal{O}(z^{3})\,,
  \\
  & \nonumber \\
  g_{E\,\rho\rho}
  & =
    \alpha'\frac{y_{\rho\rho}^{(-2)}}{z^{2}} + y_{\rho\rho}^{(0)}+
    \alpha'y_{\rho\rho}^{(0,\log)}\log{z} + \mathcal{O}(z)\,,
  \\
  & \nonumber \\
  g_{E\,\theta\theta}
  & =
    y_{\theta\theta}^{(0)}  + \mathcal{O}(z)\,,
\end{align}
\end{subequations}

\noindent
and

\begin{subequations}
\begin{align}
  F_{\rho t}
  & =
    \alpha' y^{(0)}_{F} + \mathcal{O}\left(z\right) \,,
  \\
  & \nonumber \\
  G^{(1)}{}_{\rho t}
  & =
    \alpha' y^{(0)}_{G} + \mathcal{O}\left(z\right) \,,
  \\
  & \nonumber \\
  H^{(1)}{}_{\psi \theta\varphi}
  & =
    y^{(0)}_{H} + \mathcal{O}\left(z\right) \,,
\end{align}
\end{subequations}

\noindent
where the $y_{i}$ are combinations of $d_{i}^{(2)}$, $d_{hi}^{(2)}$,
$d_{wj}^{(2)}$ and $\delta \rho$. The $\alpha'$ factors indicate the terms
that are purely first order corrections. In order to have a regular horizon we
ask that the scalars have a finite near-horizon limit, obtaining the
conditions

\begin{equation}
  y_{\phi}^{(0,\log)} = 0 \,,
  \qquad
  y_{k}^{(0,\log)} = 0 \,.
\end{equation} 

The requirement that there is an event horizon at $z = 0$ implies the
vanishing of the constant part of the $g_{E\,tt}$ component of the metric,
that is

\begin{equation}
  y_{tt}^{(0)} = 0 \,.
\end{equation}

Demanding the regularity of $g_{E\, \rho\rho}$ in the near-horizon limit (as
in the $0^{\rm th}$-order solution) we obtain

\begin{equation}
  y_{\rho\rho}^{(-2)} = 0\,,
  \qquad
  y_{\rho\rho}^{(0,\log)} = 0\,.
\end{equation}

Finally, in order to have a finite Hawking temperature, we also have to impose

\begin{equation}
  y_{tt}^{(1)} = 0\,,
  \qquad
  y_{tt}^{(2,\log)} = 0 \,.
\end{equation}
      
Combining all these conditions we obtain the following conditions for
$\delta \rho$, $d_{0}^{(2)}$, $d_{tt}^{(2)}$, $d_{\rho\rho}^{(2)}$, $d_{wj}^{(2)}$ 

\begin{subequations}
\begin{align}
  \delta \rho
  & =
    0\,,
  \\
  & \nonumber \\
 	\begin{split}
          d_{0}^{(2)}
          & =
          -\frac{\omega \left(8 q_{0}^{2}+2 q_{0} \omega+3
              \omega^{2}\right)}{8 q_{0}^{2} (2 q_{0}-\omega)} + \frac{2
            (q_{0}-\omega)}{2 q_{0}-\omega}\,d_{h0}^{(2)}
          \\
          & \\
          & \hspace{.5cm}
          +\frac{ q_{0} (2 q_{-}-\omega) (\omega-2 q_{+})}{\omega (2
            q_{0}-\omega) (q_{-}-q_{+})}\,d_{-}^{(2)}-\frac{2 q_{0}
            (q_{-}-\omega) (\omega-2 q_{+})}{\omega (2 q_{0}-\omega)
            (q_{-}-q_{+})}\,d_{h-}^{(2)}
          \\
          & \\
          & \hspace{.5cm}
 +\frac{q_{0} (2 q_{-}-\omega)}{(2 q_{0}-\omega)(q_{-}-q_{+})}\,d_{+}^{(2)}\,,
\end{split}
  \\
  & \nonumber \\
  d_{tt}^{(2)}
  & =
    \frac{\omega\left(d_{-}^{(2)} - 2 d_{h-}^{(2)} +
    d_{+}^{(2)}\right)-2q_{-}\left(d_{-}^{(2)}-d_{h-}^{(2)}\right)}{q_{+}-q_{-}}
    \,,
  \\
  & \nonumber \\
 	\begin{split}
          d_{\rho\rho}^{(2)}
          & =
          -\frac{\omega}{q_{0}}+\frac{(2 q_{-}-\omega) (-2 q_{+}+\omega)
            d_{-}^{(2)}}{(q_{-}-q_{+}) \omega}
          \\
          & \\
          & \hspace{.5cm}
          -\frac{2 (q_{-}-\omega) (-2 q_{+}+\omega) d_{h-}^{(2)}}{(q_{-}-q_{+}) \omega}+\frac{(2 q_{-}-\omega) d_{+}^{(2)}}{q_{-}-q_{+}} \,.
 	\end{split}
\end{align}
\end{subequations}

The solution still has 4 undetermined parameters: $d_{h0}^{(2)}$,
$d_{h-}^{(2)}$, $d_{+}^{(2)}$ and $d_{-}^{(2)}$ that can be determined by
demanding that the 4 independent physical charges which describe the black
hole are not modified by $\alpha'$ corrections.  Imposing these additional
conditions we get as our final result

\begin{subequations}
\begin{align}
  d_{0}^{(2)}
  & =
    -\frac{\omega \left(8 q_{0}^{2}+2 q_{0} \omega+3 \omega^{2}\right)}{8 q_{0}^{2} (2 q_{0}-\omega)} - \frac{q_{0} (2q_{-} - \omega)(2q_{+} - \omega)}{ \omega (q_{-} - q_{+})(2q_{0} - \omega)}\,d_{-}^{(2)}  \,,
  \\
  & \nonumber \\
  d^{(2)}_{tt}
  & = \frac{2q_{-} - \omega}{q_{-} - q_{+}}\,d_{-}^{(2)} \,,
  \\
  & \nonumber \\
  d_{\rho\rho}^{(2)}
  & =
    -\frac{\omega}{q_{0}} - \frac{(2q_{-} - \omega)(2q_{+} - \omega)}{\omega(q_{-}-q_{+})}\, d_{-}^{(2)} \,,
  \\
  & \nonumber \\
  d_{-}^{(2)}
  & =
    -\frac{3 \omega^{2} (q_{-} - q_{+})\left(2q_{0}
    +\omega\right)}{4 q_{0}^{2} \left[4 q_{-} q_{+} + 4 q_{0}
    (q_{+}+ q_{-})-4\left(q_{0}+q_{+}+q_{-}\right)\omega +
    3\omega^{2}\right]} \,,
  \\
  & \nonumber \\
  \delta \rho
  & =
    d_{h0}^{(2)} =  d_{h-}^{(2)} = d_{+}^{(2)} = 0 \,.
\end{align}  
\end{subequations}

\subsubsection{Horizon regularity for $\omega < 0$}

In this case, the the horizon is at $\rho = \sqrt{-\omega}$ at $0^{\rm th}$
order. At first order it could be shifted by $\alpha' \delta \rho$. We
consider therefore an expansion in $z = (\rho-\rho_{H})$ with
$\rho_{H} = \sqrt{-\omega}+\alpha' \delta \rho$. We obtain

\begin{subequations}
\begin{align}
  e^{-2\phi}
  & =
    y_{\phi}^{(0)} + \alpha' y_{\phi}^{(0,\log)} \log{z} +\mathcal{O}(z)\,,
  \\
  & \nonumber \\
  k
  & =
    y_{k}^{(0)} +  \alpha' y_{k}^{(0,\log)} \log{z} +\mathcal{O}(z)\,,
  \\
  & \nonumber \\
  g_{E\,tt}
  & =
    \alpha'y_{tt}^{(0)}+  y_{tt}^{(1)}z + \alpha'y_{tt}^{(1,\text{log})} z
    \log{z}
    +\mathcal{O}(z^{2})\,,
  \\
  & \nonumber \\
  g_{E\,\rho\rho}
  & =
    e^{-\frac{4}{3}\phi} \bigg[ \alpha'\frac{y_{\rho\rho}^{(-2)}}{z^{2}}
    +\frac{y_{\rho\rho}^{(-1)}}{z} + y_{\rho\rho}^{(0)}
    +\alpha\frac{'y_{\rho\rho}^{(-1,\log)}}{z}\log{z}+
    \alpha'y_{\rho\rho}^{(0,\log)}\log{z} +  \mathcal{O}(z) \bigg]\,,
  \\
  & \nonumber \\
  g_{E\,\theta\theta}
  & =
    y_{\theta\theta}^{(0)}  + \mathcal{O}(z)\,,
\end{align}
\end{subequations}

\noindent
and

\begin{subequations}
\begin{align}
  F_{\rho t}
  & =
    y^{(0)}_{F} +\mathcal{O}(z)\,,
  \\
  & \nonumber \\
  G^{(1)}{}_{\rho t}
  & =
    y^{(0)}_{G} +\mathcal{O}(z)\,,
  \\
  & \nonumber \\
  H^{(1)}{}_{\psi\theta\varphi}
  & =
    y^{(0)}_{H} + \mathcal{O}\left(z\right)\,,
\end{align}
\end{subequations}

\noindent
where the constants $y_{i}$ are combinations of $d_{i}$, $d_{hi}$, $d_{wj}$
and $\delta \rho$. The $\alpha'$ factors indicate the terms that are purely
first order corrections. In order to have a regular horizon we ask that the
scalars have a finite near-horizon limit. This leads to the conditions

\begin{equation}
  y_{\phi}^{(0,\log)} = 0 \,,
  \qquad
  y_{k}^{(0,\log)} = 0 \,.
\end{equation} 

The requirement that there is an event horizon at $z = 0$ implies the
vanishing of the constant part of the $g_{E\,tt}$ component of the metric

\begin{equation}
  y_{tt}^{(0)} = 0 \,.
\end{equation}

Demanding $g_{E\,\rho\rho}$ approaches the horizon at most as $1/z$ we obtain
the conditions

\begin{equation}
  y_{\rho\rho}^{(-2)} = 0\,,
  \qquad y_{\rho\rho}^{(-1,\log)} = 0\,.
\end{equation}

Finally, in order to have a finite Hawking temperature we have to impose

\begin{equation}
  y_{tt}^{(1,\log)} = 0 \,.
\end{equation}

Combining all these conditions we obtain the expressions of $\delta \rho$,
$d_{0}^{(2)}$ and $d_{-}^{(2)}$ in terms of the 5 undetermined parameters
$d_{tt}^{(2)}$, $d_{\rho\rho}^{(2)}$, $d_{h0}^{(2)}$, $d_{h-}^{(2)}$ and
$d_{+}^{(2)}$. As in the previous case, 4 integration constants can be
determined by demanding that the mass and the 3 asymptotic charges do not get
$\alpha'$ corrections. The remaining integration constant can be interpreted
as the freedom of choosing the position of the horizon, \textit{i.e.}~the
value of $\delta \rho$, and it can be eliminated through a change of
coordinates.  We finally obtain

\begin{subequations}
\begin{align}
  d_{0}^{(2)}
  & =
    \frac{\omega \left(8 q_{0}^{2}-18 q_{0} \omega+13 \omega^{2}\right)}{8
    (q_{0}-\omega)^{2} (2 q_{0}-\omega)} +
    \frac{(q_{0}-\omega)(2q_{-}-\omega)(2q_{+} - \omega)}{\omega (q_{-} -
    q_{+})(2 q_{0} - \omega)}\, d_{-}^{(2)}\,,
  \\
  & \nonumber \\
  d_{tt}^{(2)}
  & =
    \frac{2q_{-}-\omega}{q_{-} - q_{+}}\, d_{-}^{(2)} \,,
  \\
  & \nonumber \\
  d_{\rho\rho}^{(2)}
  & = 
    \frac{\omega}{q_{0}-\omega} + \frac{(2 q_{-} - \omega)(2q_{+} -
    \omega)}{\omega(q_{-}-q_{+})}\,d_{-}^{(2)} \,,
  \\
  & \nonumber \\
  d_{-}^{(2)}
  & =
    -\frac{3\omega^{2}(q_{-} -
    q_{+})(2q_{0}-3\omega)}{4(q_{0}-\omega)^{2}\left[4q_{-}q_{+}
    +4q_{0}(q_{-}+q_{+}-\omega)-4\omega\left(q_{-}+q_{+}\right)+3\omega^{2}\right]}\,,
  \\
    & \nonumber \\
  \delta \rho
  & =
    d_{h0}^{(2)} = d_{h-}^{(2)} = d_{+}^{(2)} = 0 \,.
\end{align}
\end{subequations}

\subsection{Regular Solutions}

The results of the previous section can be summarized by giving the
expressions of the corrections of the functions which lead to regular black
holes for the $\omega >0$ and $\omega<0$ cases.

\subsubsection{$\omega > 0$}

In this case, we have

\begin{subequations}
\begin{align}
  \delta \mathcal{Z}_{h0}
  & =
    \frac{2 q_{0}^{3} + \omega \left(q_{0}^{2} + 9 q_{0} \rho^{2} + 6 \rho^{4}\right)}{2q_{0} \rho^{2} (q_{0} + \rho^{2})^{2}}
    -\frac{3 \omega}{q_{0}^{2}}\log{\mathcal{Z}_{0}}\,,
  \\
  & \nonumber \\
  \begin{split}
                  \delta \mathcal{Z}_{0}
                  & =
                  \frac{8q_{0}^{6} -24q_{0}^{5}\omega +2q_{0}^{4}\omega
                    \left(7\omega -4\rho^{2}\right) +q_{0}^{3}\omega
                    \left(-4\rho^{4} -2\rho^{2}\omega
                      +\omega^{2}\right)}{4q_{0}^{2}\rho^{2}
                    \left(q_{0} +\rho^{2}\right)^{2}
                    \left(2q_{0}^{2} -3q_{0}\omega +\omega^{2}\right)}
                  \\
        	  & \\
                  & \hspace{.5cm}
                  + \frac{q_{0}^{2}\rho^{2}\omega^{2}\left(2\rho^{2}
                        +9\omega\right) +q_{0}\rho^{4}\omega^{2}
                      \left(2\rho^{2} +3\omega\right)
                      -\rho^{6}\omega^{3}}{4q_{0}^{2}\rho^{2}
                      \left(q_{0}+\rho^{2}\right)^{2}
                      \left(2q_{0}^{2} -3q_{0}\omega +\omega^{2}\right)}
                  \\
                    & \\
                    & \hspace{.5cm}
                    - \frac{q_{0}(2q_{-} -\omega)(2q_{+} -\omega)}{ \omega
                      (q_{-} -q_{+})(2q_{0} -\omega)\rho^{2} }\,d_{-}^{(2)}
                      +\frac{\omega^{2} q_{0}\left(2\rho^{2}+\omega\right)
                        -\rho^{2}\omega^{2}\left(\rho^{2} +2\omega\right)
                              }{4q_{0}^{2}\rho^{2} (q_{0}-\omega)^{2}}
                              \log{\left(\mathcal{Z}_{0}/W\right)} \,, 
                            \end{split}
  \\
  & \nonumber \\
  \delta\mathcal{Z}_{h-}
  & =
    \frac{\omega q_{-}}{2 q_{0}\rho^{4}}
    +\frac{q_{-}^{2}(2q_{+} -\omega)}{\omega(q_{-}-q_{+})\rho^{4}}\,d_{-}^{(2)}\,,  
  \\
  & \nonumber \\
  \delta\mathcal{Z}_{-}
  & =
    \delta Z_{h-} + \mathcal{Z}_{-}\left[ \Delta_{k} -
    \frac{\Delta_{C}}{\beta_{+}}+ \frac{\rho^{3}}{4}\bigg(
    \frac{\delta Z_{h-}'}{q_{-}}
    -\frac{\text{T}\left[\delta Z_{h-}'\right]}{q_{+}}\bigg)\right] \,, 
  \\
  & \nonumber \\
  \delta\mathcal{Z}_{+}
  & =
    -\mathcal{Z}_{+}\frac{\Delta_{C}}{\beta_{+}}+\text{T}\left[\delta Z_{h-}\right] \,, 
  \\
  & \nonumber \\
  \begin{split}
    \delta W_{tt}
    & =
    \frac{\omega^{2}}{2 q_{0} \rho^{4}}
    -\frac{\beta_{-}(W+\beta_{+}\beta_{-})+ W(\beta_{+} + \beta_{-})}{8
      \beta_{+} \mathcal{Z}_{0} \mathcal{Z}_{-}} \mathcal{Z}_{-}' W'
    \\
    & \\
    & \hspace{.5cm}
    +\frac{(2 q_{+} +\rho^{2})(2 q_{-} -\omega)}{(q_{-} -q_{+})\rho^{4}}d_{-}^{(2)}\,,
  \end{split}
  \\
  & \nonumber \\
  \begin{split}
    \delta W_{\rho\rho}
    & =
    \frac{\omega \left[-4 q_{0}^{3}+q_{0}^{2} \left(2 \omega-4
          \rho^{2}\right)+q_{0} \omega \left(3 \rho^{2}+\omega\right)-2
        \rho^{2} \omega \left(\rho^{2}+\omega\right)\right]}{4 q_{0}^{2}
      \rho^{2} \left(q_{0}+\rho^{2}\right) (q_{0}-\omega)}
    \\
    & \\
    & \hspace{.5cm}
    -\frac{(2q_{-}-\omega)(2q_{+}-\omega)}{\omega(q_{-}-q_{+})\rho^{2}}d_{-}^{(2)} +\frac{\omega^{2} \left(2 \rho^{4}+3 \rho^{2} \omega+\omega^{2}\right)}{4 q_{0}^{2} \rho^{2} (q_{0}-\omega)^{2}}  \log{\left(\mathcal{Z}_{0}/W\right)}\,,
        	\end{split} 
\end{align}
\end{subequations}

\noindent
where T is the operator which implements T-duality through the following
transformation of the parameters

\begin{equation}
  \label{eq:Tdualityparameters}
  q_{\pm} \leftrightarrow q_{\mp} \,,
  \qquad
  \beta_{\pm} \leftrightarrow \beta_{\mp} \,,
  \qquad
  k_{\infty} \leftrightarrow 1/k_{\infty}\,.
\end{equation}

It is easy to verify that under such transformation the solution is self-dual,
as expected.

\subsubsection{$\omega < 0$}

In this case, the corrections to the functions are given by 

\begin{subequations}
\begin{align}
  \delta \mathcal{Z}_{h0}
  & =
    \frac{2 q_{0}^{3} + \omega \left(q_{0}^{2} + 9 q_{0} \rho^{2} + 6
    \rho^{4}\right)}{2 q_{0} \rho^{2} (q_{0} + \rho^{2})^{2}}
    -\frac{3\omega}{q_{0}^{2}}
    \log{\mathcal{Z}_{0}}\,,
  \\
  & \nonumber \\
	\begin{split}
          \delta \mathcal{Z}_{0}
          & =
          \frac{8q_{0}^{6} -24q_{0}^{5}\omega -\rho^{4}\omega^{3}
            \left(\rho^{2} +2\omega\right) +q_{0}^{3}\omega
            \left(4\rho^{4} -26\rho^{2}\omega
              -7\omega^{2}\right)}{4q_{0}\rho^{2}
            \left(q_{0} +\rho^{2}\right)^{2} (q_{0} -\omega)^{2} (2 q_{0}
            -\omega)}
          \\
          & \\
          & \hspace{.5cm}
          +\frac{ \omega q_{0}^{3} \left(8\rho^{2} +22\omega\right)
            +\rho^{2}\omega^{2} \left(2\rho^{4} +11\rho^{2}\omega
              -4\omega^{2}\right)
            +q_{0}\omega^{2}
            \left(-10\rho^{4} +25\rho^{2}\omega +2\omega^{2}\right)}{4\rho^{2}
            \left(q_{0} +\rho^{2}\right)^{2} (q_{0} -\omega)^{2} (2q_{0}
            -\omega)}
          \\
          & \\
          & \hspace{.5cm}
          +\frac{(q_{0} -\omega)(2q_{-} -\omega)(2q_{+} -\omega)}{\omega(q_{-}
            -q_{+}) (2q_{0} -\omega)\rho^{2}}d_{-}^{(2)}
          +\frac{\omega^{2}q_{0}\left(2\rho^{2} +\omega\right)
            -\omega^{2}\rho^{2}\left(\rho^{2}
              +2\omega\right)}{4q_{0}^{2}\rho^{2}
            (q_{0} -\omega)^{2}} \log{\mathcal{Z}_{0}}\,, 
	\end{split}
  \\
  & \nonumber \\
          \delta\mathcal{Z}_{h-}
          & =
          -\frac{\omega q_{-}}{2 \left(q_{0} -\omega\right) \rho^{4}}
          -\frac{q_{-}(q_{-} -\omega)(2q_{+} -\omega)}{\omega (q_{-} -q_{+})
            \rho^{4}}\,d_{-}^{(2)}  \,,  
  \\
  & \nonumber \\
            \delta\mathcal{Z}_{-}
          & = 
          \delta Z_{h-} + \mathcal{Z}_{-}\left[ \Delta_{k}
            -\frac{\Delta_{C}}{\beta_{+}}
            +\frac{\rho^{3}}{4}
            \bigg( \frac{\delta Z_{h-}'}{q_{-}}
            -\frac{\text{T}\left[\delta Z_{h-}'\right]}{q_{+}}\bigg)\right]\,, 
  \\
  & \nonumber \\
  \delta\mathcal{Z}_{+}
  & =
    -\mathcal{Z}_{+}\frac{\Delta_{C}}{\beta_{+}}
    +\text{T}\left[\delta Z_{h-}\right] \,, 
  \\
  & \nonumber \\
	\begin{split}
          \delta W_{tt}
          & =
          -\frac{\omega^{2}}{2 \left(q_{0}-\omega\right) \rho^{4}}  -
          \frac{\beta_{-}(W+\beta_{+}\beta_{-})+ W(\beta_{+} + \beta_{-})}{8
            \beta_{+} \mathcal{Z}_{0} \mathcal{Z}_{-}} \mathcal{Z}_{-}' W'
          \\
          & \\
          & \hspace{.5cm}
          +\frac{(2q_{-} -\omega)(\rho^{2} +\omega)}{(q_{-} -q_{+})\rho^{4}}\,
          d_{-}^{(2)} \,,
	\end{split}
  \\
  & \nonumber \\
	\begin{split}
          \delta W_{\rho\rho}
          & =
          -\frac{\omega \left(\rho^{2} +\omega\right)\left[-4q_{0}^{3}
              +q_{0}\omega \left(5\rho^{2} -2\omega\right) +\rho^{2}\omega
              (2\rho^{2} +\omega) + q_{0}^{2} (-4 \rho^{2} +6
              \omega)\right]}{4q_{0}\rho^{4}\left(q_{0} +\rho^{2}\right)
            (q_{0} -\omega)^{2}}
          \\
          & \\
          & \hspace{.5cm}
          +\frac{(2q_{-} -\omega)(2q_{+} -\omega)(\rho^{2}
            +\omega)}{\omega(q_{-} -q_{+})\rho^{4}}d_{-}^{(2)}
          +\frac{\omega^{2} \left(2 \rho^{4} +3 \rho^{2}\omega
              +\omega^{2}\right)}{4 q_{0}^{2} \rho^{2} (q_{0} -\omega)^{2}}
          \log{\mathcal{Z}_{0}} \,,
	\end{split} 
\end{align}
\end{subequations}

\noindent
where, again, T is the operator that implements T~duality as in
Eqs.~(\ref{eq:Tdualityparameters}).

As we have discussed in Appendix~\ref{app-zerothorder}, at $0^{\rm th}$ order the
$\omega > 0$ and $\omega < 0$ solutions are related by the coordinate
transformation Eq.~(\ref{eq:omegatominusomegacoordinatetransformation})
supplemented by the transformations of the parameters
Eqs.~(\ref{eq:omegatominusomega0}) and (\ref{eq:omegatominusomega}).  This is
still true at first order in $\alpha'$, as can be seen by checking the
transformations of all the physical fields. However, in the $0^{\rm th}$-order case,
the two solutions have the same form and the metric is formally invariant
under those transformations plus $\omega\rightarrow -\omega$, instead in the
first-order case the two solutions have different forms and there is no
invariance under those transformations. Thus, the mass and asymptotic charges
do not need to be invariant under them, as in the $0^{\rm th}$-order
case. Nevertheless they do happen to be invariant.  On the other hand, the
realization of these transformations at the level of the $\mathcal{Z}$s and $W$s
functions is far from evident.

\subsubsection{$\omega = 0$}

As a further check of the solutions we have obtained, we can 
take the $|\omega| \rightarrow 0$ limit

\begin{subequations}
\begin{align}
  \delta \mathcal{Z}_{h0}
  & =
    \delta \mathcal{Z}_{0}
    =
    \frac{q_{0}^{2}}{\rho^{2}(q_{0}+\rho^{2})^{2}}\,,
  \\
  & \nonumber \\
  \delta \mathcal{Z}_{+}
  & =
    - (1 + s_{+}s_{-})\frac{q_{+}q_{-}}{\rho^{2}(q_{0}+\rho^{2})(q_{-}+\rho^{2})} \,,
  \\
  & \nonumber \\
  \delta \mathcal{Z}_{h-}
  & =
    \delta \mathcal{Z}_{-}
    =
    \delta W_{tt} = \delta W_{\rho\rho} = 0\,.
\end{align}
\end{subequations}

\noindent
and compare with the extremal $\alpha'$-corrected solutions of
Ref.~\cite{Cano:2021nzo}.\footnote{It has to be taken into account that the
  $q_{i}$ parameters we are using here are equivalent to the $\hat{q}_{i}$s in
  Ref.~\cite{Cano:2021nzo}.}

\subsubsection{The Reissner-Nordstr\"om-Tangherlini limit}

This is the $q_{0}=q_{+}=q_{-}=q$ case reviewed at $0^{\rm th}$ order in
Appendix~\ref{app-RNT}. Using the $\omega>0$ solution, we get

\begin{subequations}
  \begin{align}
    \delta \mathcal{Z}_{h0}
    & =
      \frac{2 q^{3}+\omega \left(q^{2}+9 q \rho^{2}+6
      \rho^{4}\right)}{2 q \rho^{2}
      \left(q+\rho^{2}\right)^{2}}
      -\frac{3 \omega}{q^{2}} \log{\mathcal{Z}}\,,
    \\
    & \nonumber \\
    \begin{split}
      \delta \mathcal{Z}_{0}
      & =
      \frac{8 q^{6}-22 q^{5} \omega+q^{4} \omega
        \left(13 \omega-4 \rho^{2}\right)-2 q^{3} \rho^{2} \omega
        \left(\rho^{2}+2 \omega\right)}{4 q^{2} \rho^{2}
        \left(q+\rho^{2}\right)^{2} \left(2 q^{2}-3 q
          \omega+\omega^{2}\right)}
      \\
      & \\
      & \hspace{.5cm} +\frac{q^{2} \rho^{2} \omega^{2} \left(\rho^{2}+7
          \omega\right)+2 q \rho^{4} \omega^{2}
        \left(\rho^{2}+\omega\right)-\rho^{6} \omega^{3}}{4 q^{2} \rho^{2}
        \left(q+\rho^{2}\right)^{2} \left(2 q^{2}-3 q
          \omega+\omega^{2}\right)}
      \\
      & \\
      & \hspace{.5cm} +\frac{\omega^{2} \left[q \left(2
            \rho^{2}+\omega\right)-\rho^{2} \left(\rho^{2}+2
            \omega\right)\right]}{4 q^{2} \rho^{2} (q-\omega)^{2}}
      \log{\left(\mathcal{Z}/W\right)}\,,
		\end{split}
    \\
    & \nonumber \\
    \delta\mathcal{Z}_{h-}
    & =
      \frac{\omega (2 q-3 \omega)}{4 \rho^{4} (2 q-\omega)} \,,  
    \\
    & \nonumber \\
      \delta\mathcal{Z}_{-}
      & =
      \frac{\omega (2 q-3 \omega)}{4 \rho^{4} (2q-\omega)}
      -(1+s_{+}s_{-})\frac{q \omega}{2 \rho^{4}\left(q+\rho^{2}\right)} \,,
    \\
    & \nonumber \\
    \delta\mathcal{Z}_{+}
    & =
      \frac{\omega (2 q-3 \omega)}{4 \rho^{4} (2 q-\omega)}
      +(1+s_+s_-)\frac{ \rho^{2} q \omega-q^{2} \left(2 \rho^{2}+\omega\right)}{2 \rho^{4} \left(q+\rho^{2}\right)^{2}} \,, 
    \\
    & \nonumber \\
    \delta W_{tt}
    & =
      \frac{\omega^{2} \left(2 q+\rho^{2}\right)
      \left(2q^{2}-2q\rho^{2}-3q \omega-\rho^{2} \omega\right)}{4 q^{2} \rho^{4}
      \left(q+\rho^{2}\right) (2 q-\omega)}
      -(1+s_+s_-)\frac{q \omega \left(\rho^{2}+\omega\right)}{\rho^{4}
      \left(\rho^{2}+q\right)^{2}}\,,
    \\
    & \nonumber \\
    \begin{split}
      \delta W_{\rho\rho}
      & =
      -\frac{\omega \left[2 q^{3}+q^{2} \left(2
            \rho^{2}-\omega\right)-2 q \rho^{2} \omega+\rho^{2} \omega \left(2
            \rho^{2}+3 \omega\right)\right]}{4 q^{2} \rho^{2}
        \left(q+\rho^{2}\right) (q-\omega)}
      \\
      & \\
      & \hspace{.5cm}
      +\frac{\omega^{2} \left(2\rho^{4} +3\rho^{2}\omega
          +\omega^{2}\right)}{4q^{2}\rho^{2}(q-\omega)^{2}}
      \log{\left(\mathcal{Z}/W\right)}\,.
		\end{split} 
	\end{align}
\end{subequations}

\begin{figure}[t!]
\centering
\includegraphics[width=0.7\textwidth]{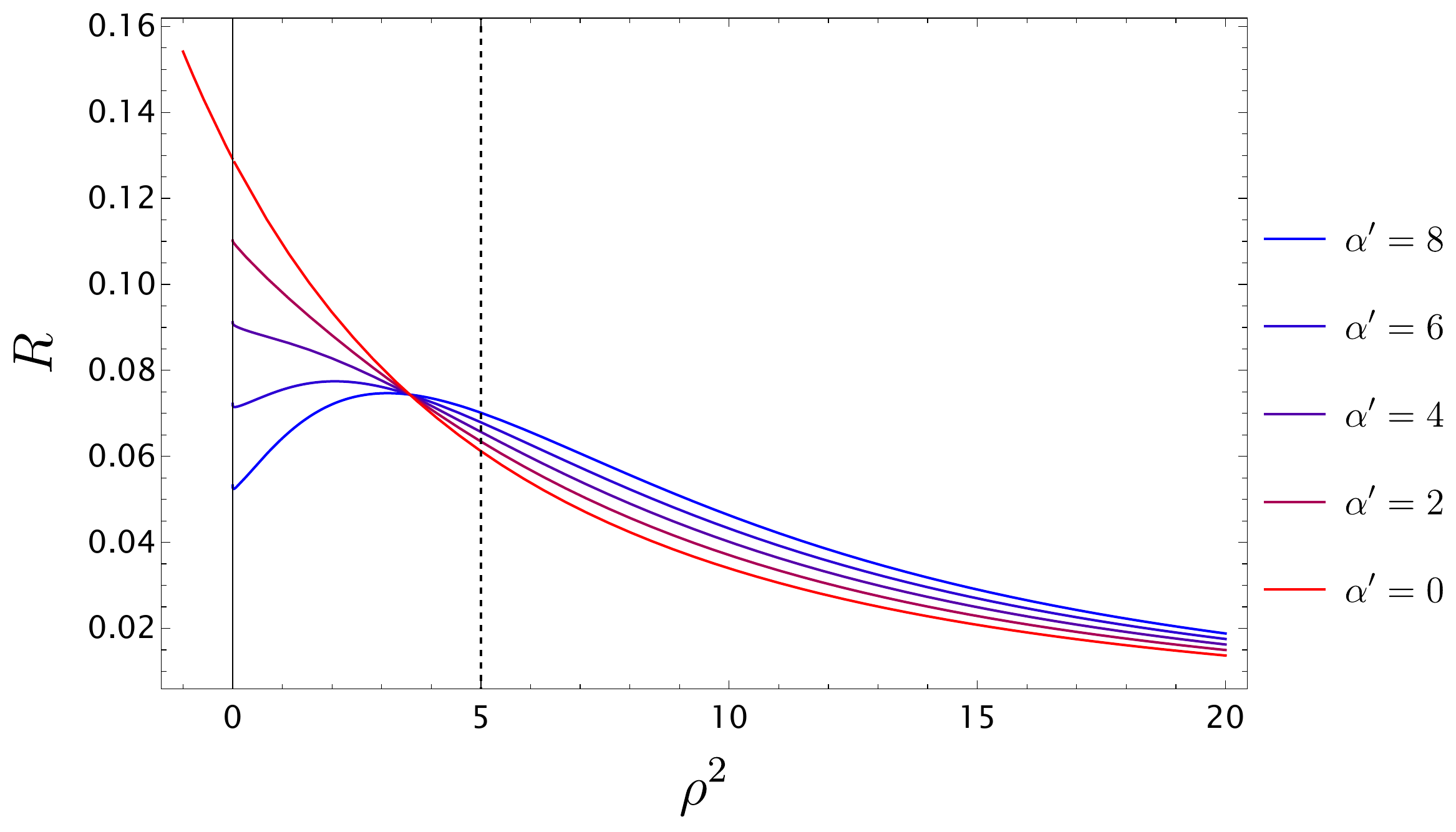}
\caption{\small the Ricci scalar as a function of the radial coordinate for
  $q_{+} = 40$, $q_{-} = 20$, $q_{0} = 10$, $\omega = -5$, $s_{+}s_{-} = -1$,
  for different values of $\alpha'$. Observe that, for these charges, the
  black hole does not become supersymmetric in the extremal limit.}
\label{fig:P1}
\end{figure}

\begin{figure}[t!]
  \centering \includegraphics[width=0.7\textwidth]{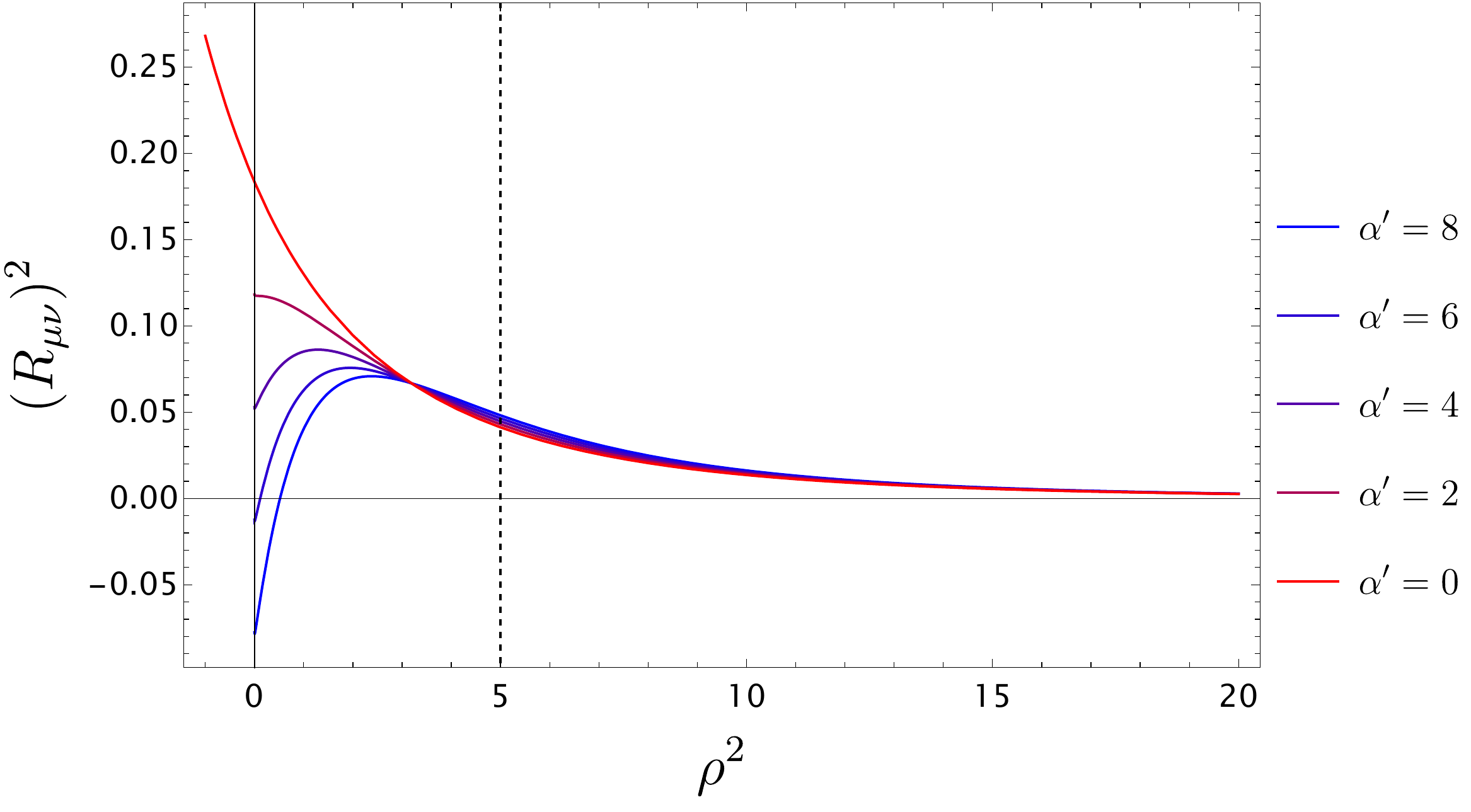}
  \caption{\small The $R_{\mu\nu}R^{\mu\nu}$ invariant as a function of the
    radial coordinate for $q_{+} = 40$, $q_{-} = 20$, $q_{0} = 10$,
    $\omega = -5$, $s_{+}s_{-} = -1$ for different values of
    $\alpha'$. Observe that, for these charges, the black hole does not become
    supersymmetric in the extremal limit.}
\label{fig:P2}
\end{figure}

\begin{figure}[t!]
  \centering \includegraphics[width=0.7\textwidth]{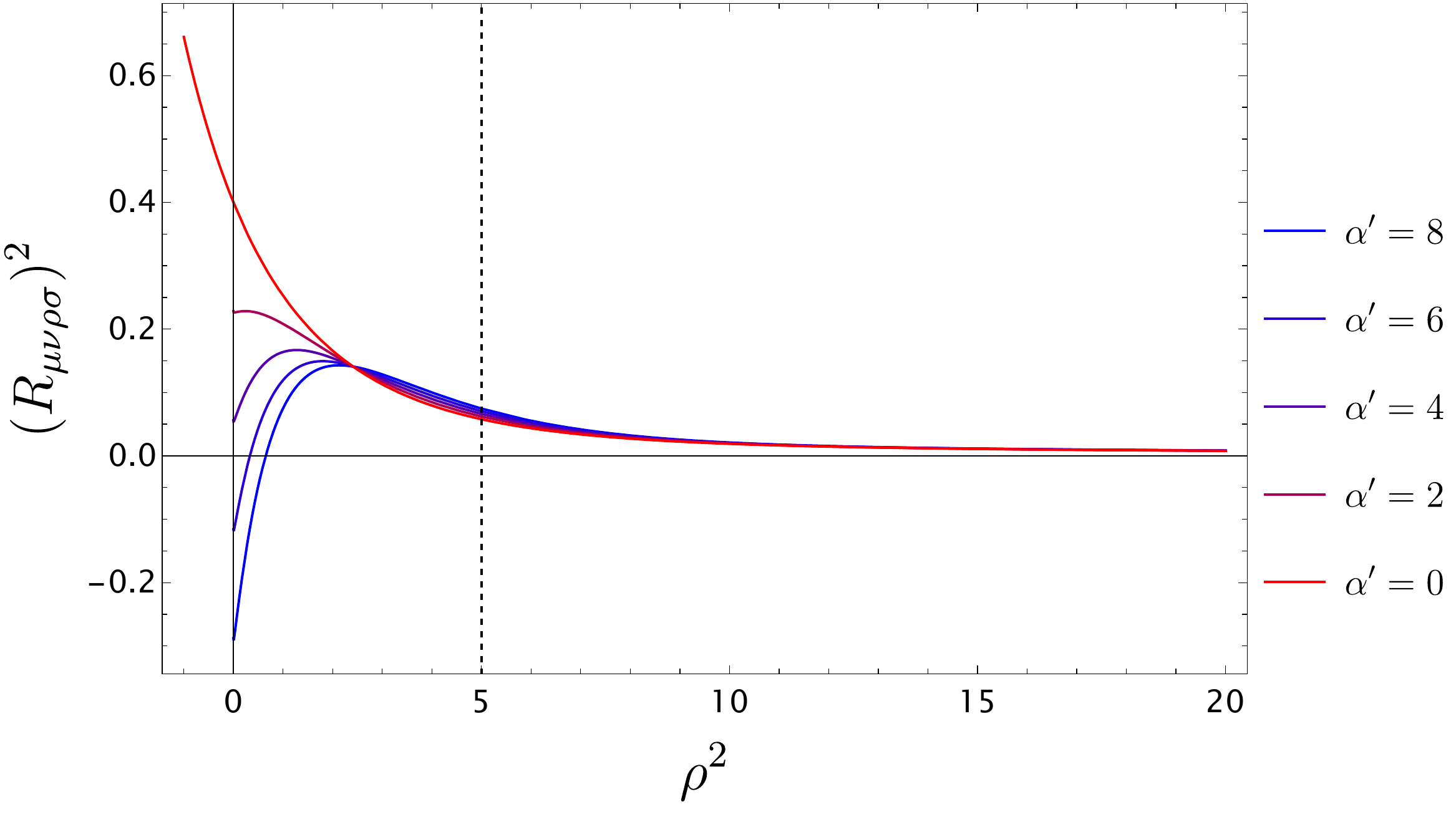}
  \caption{\small The Kretschmann invariant
    $R_{\mu\nu\rho\sigma}R^{\mu\nu\rho\sigma}$ as a function of the radial
    coordinate for $q_{+} = 40$, $q_{-} = 20$, $q_{0} = 10$, $\omega = -5$,
    $s_{+}s_{-} = -1$, for different values
  of $\alpha'$. Observe that, for these charges, the black hole does not become
    supersymmetric in the extremal limit.}
\label{fig:P3}
\end{figure}

To end this section, we present Figs.~\ref{fig:P1}, \ref{fig:P2} and
\ref{fig:P3} in which we plot several curvature invariants for a typical
choice of integration constants $q_{+},q_{-},q_{0},\omega$ with $\omega < 0$
and for several values of $\alpha'$ as a way to visualize the effect of those
corrections which have to be small anyway. The plots do not extend beyond
$\rho^{2}=0$ for $\alpha'\neq 0$ because there is a logarithmic singularity at
that point. There seem to be no other curvature singularities for larger
values of $\rho^{2}$, which include the position of the inner horizon, which
is slightly displaced to the right of $\rho^{2}=0$ by the $\alpha'$
corrections.

\section{Thermodynamics}
\label{sec-thermodynamics}

We are going to study separately the $\omega < 0$ and $\omega > 0$ cases as a
further test of our solution.

\subsection{$\omega > 0$}

The physical charges of these solutions can be computed using the following
definitions:

\begin{subequations}
  \label{eq:5dchargesdeffirstorder}
  \begin{align}
    Q^{(1)}_{+}
    & =
      \frac{-1}{16\pi G_{N}^{(5)}}
      \int_{S^{3}_{\infty}}\left\{ e^{-4\phi/3}k^{2}\star F\right\}
      =
      \frac{\pi}{4G_{N}^{(5)}}k_{\infty}\beta_{+}q_{+}\,,
    \\
    & \nonumber \\
    Q^{(1)}_{-}
    & =
      \frac{-1}{16\pi G_{N}^{(5)}}
      \int_{S^{3}_{\infty}} \left\{ e^{-4\phi/3}k^{(1)\,-2}\star G^{(1)}\right\}
      =
      \frac{\pi}{4G_{N}^{(5)}}k^{-1}_{\infty}\beta_{-}q_{-}\,,
    \\
    & \nonumber \\
    Q^{(1)}_{0}
    & =
      \frac{-1}{16\pi G_{N}^{(5)}}
      \int_{S^{3}_{\infty}} \left\{ e^{8\phi/3}\star K^{(1)}\right\}
      =
      -\frac{\pi}{4G_{N}^{(5)}}\beta_{0}q_{0}\,.
  \end{align}
\end{subequations}

These definitions are similar to those used at $0^{\rm th}$ order
Eqs.~(\ref{eq:5dchargesdef}) with the replacements of $0^{\rm th}$-order field
strengths and fields by the first-order ones, and the result is formally
identical. We have ignored the second terms in the integrands because, as we
have explained, they vanish for these solutions, but we have ignored many
higher-order terms as well because they fall off fast for large values of
$\rho$ and they do not contribute at infinity. They would give a finite
contribution elsewhere, though. Therefore, these charges are not
\textit{localized} in the language of Ref.~\cite{Marolf:2000cb} and they are
defined strictly at infinity.

In agreement with the choices made, the mass is given by the $0^{\rm th}$-order
expression Eq.~(\ref{eq:Mzerothorder}).

The horizon is at $\rho= \rho_{H}= 0$ in these coordinates. The radius of the
horizon, defined by Eq.~(\ref{eq:radiusofthehorizon}) is given at first order
($R_{H}^{(1)}$) by

\begin{equation}
  R_{H}^{(1)}
  =
  R_{H}^{(0)}
  \left[1-\alpha'\, \frac{3 \omega + 4 q_{0} \beta_{+}\beta_{-}}{24 q_{0}}\right]\,,
\end{equation}

\noindent
where the radius of the horizon of the $0^{\rm th}$-order solution, $R_{H}^{(0)}$ is
given in Eq.~(\ref{eq:RH0omegapositive}), and we obtain the first-order
Bekenstein-Hawking entropy $S_{BH}^{(1)}$ in terms of the $0^{\rm th}$-order one in
Eq.~(\ref{eq:zerothorderBHentropyomegapositive})

\begin{equation}
  S_{BH}^{(1)}
  =
  S_{BH}^{(0)}
  \left[1-\alpha'\, \frac{3 \omega + 4 q_{0} \beta_{+}\beta_{-}}{8 q_{0}}\right]\,.
\end{equation}

The Hawking temperature is now

\begin{equation}
  \begin{split}
    T^{(1)}_{H}
    & =
    T^{(0)}_{H}
    \bigg\{1+\alpha'\bigg[\frac{\beta_{-} \beta_{+}}{2 q_{0}}
    +\frac{\omega \left(8 q_{0}^{2}+q_{0} q_{-}+q_{0} q_{+}+5 q_{-} q_{+}\right)}{2q_{0}^{2}D}
    \\
    & \\
    &  \hspace{.5cm}
    +\frac{16 q_{0}(q_{-}q_{+} -2 q_{0}q_{-} -2 q_{0}q_{+})
      -8 \omega^{2} (q_{0}+q_{-}+q_{+})+3 \omega^{3}}{8q_{0}^{2}D} \bigg] \bigg\}\,,
	\end{split}
\end{equation}

\noindent
where $T_{H}^{(0)}$ is the Hawing temperature of the $0^{\rm th}$-order solution
Eq.~(\ref{eq:zerothorderHtemperatureomegapositive}) and where we have defined

\begin{equation}
  \label{eq:D}
  D
  \equiv
  4q_{-}q_{+}    -4\omega(q_{0}+q_{-}+q_{+})    +4q_{0}(q_{-}+q_{+})+3 \omega^{2}\,.
\end{equation}

\noindent
Since $T_{H}^{(1)}$ is proportional to $T_{H}^{(0)}$, it vanishes when
$T_{H}^{(0)}$ vanishes, namely in the extremal limit $\omega=0$.

We can compute the Wald entropy using the gauge-invariant formula of
Ref.~\cite{Elgood:2020nls}. We obtain

\begin{equation}
  S^{(1)}_{W}
  =
  S^{(0)}_{BH}
  \left[1+\alpha'\,\frac{8q_{0} +\omega +4q_{0}\beta_{+}\beta_{-}}{8q_{0}^{2}}\right]\,,
\end{equation}

\noindent
where $ S^{(0)}_{BH}$ is the Bekenstein-Hawking entropy of the $0^{\rm th}$-order
solution Eq.~(\ref{eq:zerothorderBHentropyomegapositive}).

Notice the relations

\begin{subequations}
  \begin{align}
    S_{BH}^{(1)}T^{(1)}_{H}
    & =
      \frac{\pi\omega}{4G_{N}^{(5)}}
      \bigg\{1+\alpha' \bigg[\frac{q_{-}q_{+}
      -2q_{0}(q_{-}+q_{+}-\omega)}{q_{0}^{2}D}(2q_{0} +\omega)
    \nonumber \\
    & \nonumber \\
    & \hspace{.5cm}
      +\frac{2\omega(q_{-}+q_{+}) -3\omega^{2}}{4q_{0}^{2}D}
      (2q_{0} +\omega)\bigg]\, \bigg\}\,,
    \\
    & \nonumber \\
    S_{W}^{(1)}T^{(1)}_{H}
    & = 
      \frac{\pi\omega}{4G_{N}^{(5)}}
      \bigg\{1+\alpha'\bigg[\frac{\beta_{-}\beta_{+}}{q_{0}}
      +\frac{3 (2 q_{-}-\omega) (2 q_{+}-\omega)
      (2q_{0} +\omega)}{4 q_0^2 D}\bigg]\bigg\}\,.
  \end{align}
\end{subequations}

Using the $\omega >0$ form of the solution we cannot compute these products on
the inner horizon and check the property Eq.~(\ref{eq:TSproductszerothorder})
because the inner horizon is not covered by the coordinates we are using.  We
can, however, use the latter to check the Smarr formula. Using the general
definitions Eqs.~(\ref{eq:definitionselectrostaticpotentials}), we find that,
at first order in $\alpha'$, the electrostatic potentials evaluated on the
horizon are given by

\begin{subequations}
  \label{eq:firstorderpotentials}
	\begin{align}
          \Phi_{+}
          & =
            k_\infty^{-1} \beta_{+} \bigg\{1 + \alpha'\bigg[-
            \frac{\omega\beta_{-}}{2 q_{0} \beta_{+} q_{+}}
            -\frac{3\omega (2 q_{0}+\omega) (2q_{-}-\omega)}{4 q_{0}^{2}
            D}\bigg]\bigg\} \,,
          \\
          & \nonumber \\
          \Phi_{-}
          & =
            k_\infty\, \beta_{-} \bigg\{1 + \alpha'\bigg[
            -\frac{\omega\beta_{+}}{2 q_{0} \beta_{-} q_{-}}
            -\frac{3\omega (2 q_{0}+\omega) (2q_{+}-\omega)}{4 q_{0}^{2}
            D}\bigg]\bigg\} \,,
          \\
          & \nonumber \\
                  \Phi_{0}
                  & =
                  -\beta_{0} \bigg\{1 + \alpha'
                  \bigg[ \frac{4q_{0}\omega
                    (q_{-}+q_{+}-\omega) +q_{-}q_{+}\omega}{ q_{0}^{2} D}
                  +\frac{2\omega^{2}( q_{+}+q_{-})
                    -3 \omega^{3}}{4 q_{0}^{2} D} \bigg] \bigg\} \,.
	\end{align}
\end{subequations}

If we now plug all this information into the $0^{\rm th}$-order Smarr formula
Eq.~(\ref{eq:zerothSmarrformula}), we find that it is only satisfied up to a
term of first order in $\alpha'$. As a matter of fact, this could have been
expected: all the dimensionful parameters in the action can be interpreted as
thermodynamical variables and come with their own conjugate thermodynamical
potentials \cite{Kastor:2010gq,Hajian:2021hje,Ortin:2021ade}. Thus, the
coefficient of $\alpha'$ in the additional term should be interpreted as the
thermodynamical potential conjugate to $\alpha'$ that we are going to denote
by $\Phi^{\alpha'}$ and which takes the value

\begin{equation}
  \label{eq:Phialphaomegapositive}
  \Phi^{\alpha'}
  =
  -\frac{\pi \omega}{4G_{N}^{(5)}}
  \frac{8q_{0}+\omega+4q_{0}\beta_{+}\beta_{-}}{8q_{0}^{2}}\,.
\end{equation}

Then, the Smarr formula takes the form

\begin{equation}
  \label{eq:firstSmarrformula}
  M^{(1)}
  =
  \tfrac{3}{2}S_{W}^{(1)}T_{H}^{(1)} +\Phi^{+}Q_{+} +\Phi^{-}Q_{-}
  +\Phi^{0}Q_{0}  +\Phi^{\alpha'}\alpha'\,.
\end{equation}

If this interpretation is correct, then we must find a term
$\Phi^{\alpha'}\delta \alpha'$ in the first law, which should take the form

\begin{equation}
  \label{eq:firstfirstlaw}
  \delta M^{(1)}
  =
  T_{H}^{(1)}\delta S_{W}^{(1)} +\Phi^{+}\delta Q_{+}
  +\Phi^{-}\delta Q_{-}  +\Phi^{0}\delta Q_{0}
  +\Phi^{\alpha'}\delta \alpha'\,.
\end{equation}

This is what we are going to check next.

\subsubsection{Checking the first law at first order in $\alpha'$}

The procedure we are going to follow to check the first law at first order is
essentially the same we followed to check it a $0^{\rm th}$ order, in
Appendix~\ref{sec-checkingfirstlawzeroth}. Now we will also consider the
variation of $\alpha'$ to test the proposal made in the previous section,
taking into account that the parameters $\beta_{i},q_{i},\omega$ are
independent of $\alpha'$.

The variations of the physical charges can be easily obtained from
Eqs.~(\ref{eq:5dchargesdeffirstorder})

\begin{equation}
  \label{eq026}
  \delta Q^{(1)}_{i}
  = -\frac{\pi}{4 G_{N}^{(5)}}\,
  (-k_\infty)^{\gamma_{i}} \, q_{i} \beta_{i}
  \left[\frac{2 q_{i} - \omega}{2q_{i}(q_{i} -\omega)}\delta q_{i}
    -\frac{1}{2(q_{i}-\omega)} \delta \omega \right]\,,
\end{equation}

\noindent
with

\begin{equation}
	\gamma_{i} = \begin{cases}
		\pm1 & \text{if}\quad i = \pm  \\
		0 & \text{if}\quad i = 0
	\end{cases}
\end{equation}

The variation of the mass is 

\begin{equation}
  \label{eq027}
  \delta M^{(1)}
  = \frac{\pi}{4 G_{N}^{(5)}}
  \left[\delta q_{-} +\delta q_{+} +\delta q_{0} \right]
  -\frac{3 \pi}{8 G_{N}^{(5)}}\delta \omega \,.
\end{equation}

The variation of the entropy is 

\begin{equation}
  \label{eq028}
	\begin{split}
          \frac{\delta S^{(1)}_{W}}{S_{H}^{(0)}}
          = &
          \left[\frac{1}{2q_{0}}+ \alpha'\left(-\frac{\beta_{-} \beta_{+}}{4
                q_{0}^{2}} -\frac{8 q_{0} + 3\omega}{16 q_{0}^{3}}
            \right)\right]\delta q_{0}
          \\
          & \\
          &
          +\left[\frac{1}{2q_{-}} + \alpha' \left(\frac{\beta_{+}}{4q_{0}
                q_{-} \beta_{-}} + \frac{8 q_{0} + \omega}{16 q_{0}^{2}
                q_{-}}\right)\right] \delta q_{-}
          \\
          & \\
          &
          +\left[\frac{1}{2q_{+}} + \alpha' \left(\frac{\beta_{-}}{4q_{0}
                q_{+} \beta_{+}} + \frac{8 q_{0} + \omega}{16 q_{0}^{2}
                q_{+}}\right)\right] \delta q_{+}
          \\
          & \\
          &
          +\left[\frac{(2\omega-q_{+}-q_{-})}{4q_{0}q_{-}q_{+}\beta_{+}\beta_{-}}
            +\frac{1}{8q_{0}^{2}}\right]\alpha' \delta \omega
                    \\
          & \\
          &
          +\left[\frac{8q_{0} +\omega +4q_{0}\beta_{+}\beta_{-}}{8q_{0}^{2}}\right]
          \delta\alpha'\,. 
	\end{split}
\end{equation}

Expressing $\delta q_{i}$ and $\delta \omega$ in terms of $\delta Q_{i}^{(1)}$
and $\delta M^{(1)}$ through Eqs.~(\ref{eq026}) and (\ref{eq027}) and
replacing them into (\ref{eq028}) we obtain after some algebra

\begin{equation}
  \delta S^{(1)}_{W}
  =
  \frac{1}{T^{(1)}_{H}} \bigg[ \delta M^{(1)}
  -\Phi_{+} \delta Q_{+}^{(1)} -\Phi_{-}\delta Q_{-}^{(1)}
  -\Phi_{0} \delta Q_{0}^{(1)}
  -\Phi^{\alpha'}\delta\alpha'\bigg]\,,
\end{equation}

\noindent
where the potentials $\Phi^{\pm,0}$ are
given by Eqs.~(\ref{eq:firstorderpotentials}) and $\Phi^{\alpha'}$ has the
expression Eq.~(\ref{eq:Phialphaomegapositive}) that we proposed to make sense
of the Smarr formula. This confirms our identification and interpretation.

\subsubsection{The Reissner-Nordstr\"om-Tangherlini case with $\omega>0$}

Due to the choices made, the relations between the
integration parameters $\omega,q,\beta$ and the physical parameters
$M,Q=Q_{0}$ ($Q_{\pm}=k_{\infty}^{\pm 1}Q_{0}$) are the $0^{\rm th}$-order ones
Eqs.~(\ref{eq:integrationparametersversusphysicalparameters}),  which, in its
turn, makes it easy to express the corrected thermodynamical quantities in
terms of the latter.

\begin{subequations}
  \begin{align}
    R_{H}^{(1)}
    & =
      R_{H}^{(0)}
      \left\{1-\frac{\alpha'}{24q^{2}}
      \left[3\omega +4s_{+}s_{-}(q-\omega)\right]\right\}\,,
    \\
    & \nonumber \\
    T_{H}^{(1)}
    & =
      T_{H}^{(0)}
      \left\{1+\frac{\alpha'}{8q^{2}}
      \left[\omega-4q +4s_{+}s_{-}(q-\omega)\right]\right\}\,,
    \\
    & \nonumber \\
    S_{BH}^{(1)}
    & =
      S_{BH}^{(0)}\left\{1-\frac{\alpha'}{8q^{2}}
      \left[3\omega +4s_{+}s_{-}(q-\omega)\right] \right\}\,,
    \\
    & \nonumber \\
    S_{W}^{(1)}
    & =
      S_{BH}^{(0)}\left\{1+\frac{\alpha'}{8q^{2}}
      \left[(\omega+8q) +4s_{+}s_{-}(q-\omega)\right] \right\}\,,
  \end{align}
\end{subequations}



\subsection{$\omega < 0$}

The thermodynamic quantities can be easily obtained from those computed in the
previous section for the $\omega>0$ case through the transformations
Eqs.~(\ref{eq:omegatominusomega0}) and (\ref{eq:omegatominusomega}).
after which we have to consider $\omega<0$.

The asymptotic charges are still given by
Eqs.~(\ref{eq:5dchargesdeffirstorder}) and the mass still takes the same form
as in the $0^{\rm th}$-order solution Eq.~(\ref{eq:Mzerothorder}).  The horizon
radius, the Hawking temperature, and the Bekenstein-Hawking and Wald entropies
are given by

\begin{subequations}
  \begin{align}
  R_{H}^{(1)}
  & =
  R_{H}^{(0)}
    \left\{1 +\alpha'\,\left[\frac{\omega}{8 (q_{0} -\omega)^{2}}
    -\frac{1}{6(q_{0} -\omega)\beta_{+}\beta_{-}}\right] \right\} \,,
    \\
    & \nonumber \\
 	\begin{split}
          T^{(1)}_{H}
          & =
          T^{(0)}_{H}
          \bigg\{1+\alpha'\bigg[\frac{1}{2(q_{0} -\omega)\beta_{-}\beta_{+}}
          -\frac{8q_{0}^{2}(q_{-}+q_{+}-\omega)+9q_{-}q_{+}\omega}{2(q_{0}-\omega)^{2}D}
          \\
          & \\
          & \hspace{.5cm}
          + \frac{4q_{0}\left(4q_{-}q_{+} +11q_{-}\omega +11q_{+}\omega
              -12\omega^{2}\right) +9\omega^{3}}{8(q_{0}-\omega)^{2}D} \bigg] \bigg\}\,,
	\end{split}
    \\
    & \nonumber \\
    S_{BH}^{(1)}
    & =
      S_{BH}^{(0)}
      \left\{1+\alpha'\,\left[\frac{3\omega}{8 (q_{0} -\omega)^{2}}
      -\frac{1}{2(q_{0} -\omega) \beta_{+}\beta_{-}}\right] \right\}\,,
    \\
    & \nonumber \\
    S^{(1)}_{W}
    & =
      S^{(0)}_{BH}
      \left\{1+\alpha'\,\left[\frac{8q_{0} - 9\omega}{8(q_{0} -\omega)^{2}}
      + \frac{1}{2(q_{0} -\omega)\beta_{+}\beta_{-}}\right] \right\} \,,
  \end{align}
\end{subequations}

\noindent
where $D$ is still given by Eq.~(\ref{eq:D}).

Notice the relations

\begin{subequations}
  \begin{align}
    S_{BH}^{(1)}T^{(1)}_{H}
    & =
      - \frac{\pi\omega}{4G_{N}^{(5)}}
      \bigg\{1 +\alpha' \bigg[
      -\frac{\left(8q_{0}q_{-} +8q_{0}q_{+} -4q_{-}q_{+} +3\omega^{2}\right)
      (2q_{0} -3\omega)}{4(q_{0}-\omega)^{2}D}
      \nonumber \\
    & \nonumber \\
    & \hspace{.5cm}
      +\frac{\omega(4q_{0} +q_{-} +q_{+})
      (2q_{0} -3\omega)}{2(q_{0}-\omega)^{2}D}\bigg]\, \bigg\}\,,
    \\
    & \nonumber \\
    S_{W}^{(1)}T^{(1)}_{H}
    & =
      -\frac{\pi\omega}{4G_{N}^{(5)}}
      \bigg\{1 +\alpha'\bigg[\frac{1}{(q_{0} -\omega)\beta_{-}\beta_{+}}
      +\frac{3(2 q_{-} -\omega) (2q_{+} -\omega)
      (2q_{0} - 3\omega)}{4(q_{0}-\omega)^{2}D}\bigg]\bigg\}\,.
  \end{align}
\end{subequations}


Now the electrostatic potential conjugate to the charges, still defined by
Eqs.~(\ref{eq:definitionselectrostaticpotentials}) are given by

\begin{subequations}
  \label{eq:potentialsforomeganegative}
	\begin{align}
          \Phi_{+}
          & =
            \frac{k_\infty^{-1}}{\beta_{+}}  \bigg\{1 + \alpha'\bigg[
            \frac{\omega\beta_{+}}{2 (q_{0}-\omega) (q_{+}-\omega) \beta_{-} }
            +\frac{3 \omega (2q_{0}-3 \omega )(2q_{-} - \omega)}{4
            (q_{0}-\omega)^{2}{D}}\bigg]\bigg\} \,,
          \\
          & \nonumber \\
          \Phi_{-}
          & =
            \frac{k_\infty}{\beta_{-}}\bigg\{1 + \alpha'\bigg[
            \frac{\omega\beta_{-}}{2 (q_{0}-\omega) (q_{-}-\omega) \beta_{+} }
            +\frac{3 \omega (2q_{0}-3 \omega )(2q_{+} - \omega)}{4
            (q_{0}-\omega)^{2}{D}}\bigg]\bigg\} \,,
          \\
          & \nonumber \\
          \Phi_{0}
          & =
            -\frac{1}{\beta_{0}} \bigg\{1 - \alpha'\bigg[ \frac{4q_{-}q_{+}\omega + 16 q_{0} \omega (q_{-}+q_{+}-\omega)- 22\omega^{2}(q_{-}+q_{+})+21 \omega^{3}}{4 (q_{0} -\omega)^{2}D} \bigg] \bigg\}\,,
	\end{align}
\end{subequations}

\noindent
and the Smarr formula Eq.~(\ref{eq:firstSmarrformula}) is satisfied with

\begin{equation}
  \label{eq:Phialphaomeganegative}
  \Phi^{\alpha'}
  =
  -\frac{\pi \omega}{4G_{N}^{(5)}}
  \frac{(8q_{0}-9\omega)\beta_{+}\beta_{-}
    +4(q_{0}-\omega)}{8(q_{0}-\omega)^{2}\beta_{+}\beta_{-}}\,.
\end{equation}

We have checked explicitly that the first law in the form
Eq.~(\ref{eq:firstfirstlaw}) is satisfied with the above potentials.

\subsubsection{The Reissner-Nordstr\"om-Tangherlini case with $\omega<0$}

Al the physical quantities can be computed by performing the transformations
Eqs.~(\ref{eq:omegatominusomega0}) and (\ref{eq:omegatominusomega}) on the
expressions we found for the $\omega>0$ case and we will not write them
explicitly here.

\begin{figure}[t!]
  \centering \includegraphics[width=0.6\textwidth]{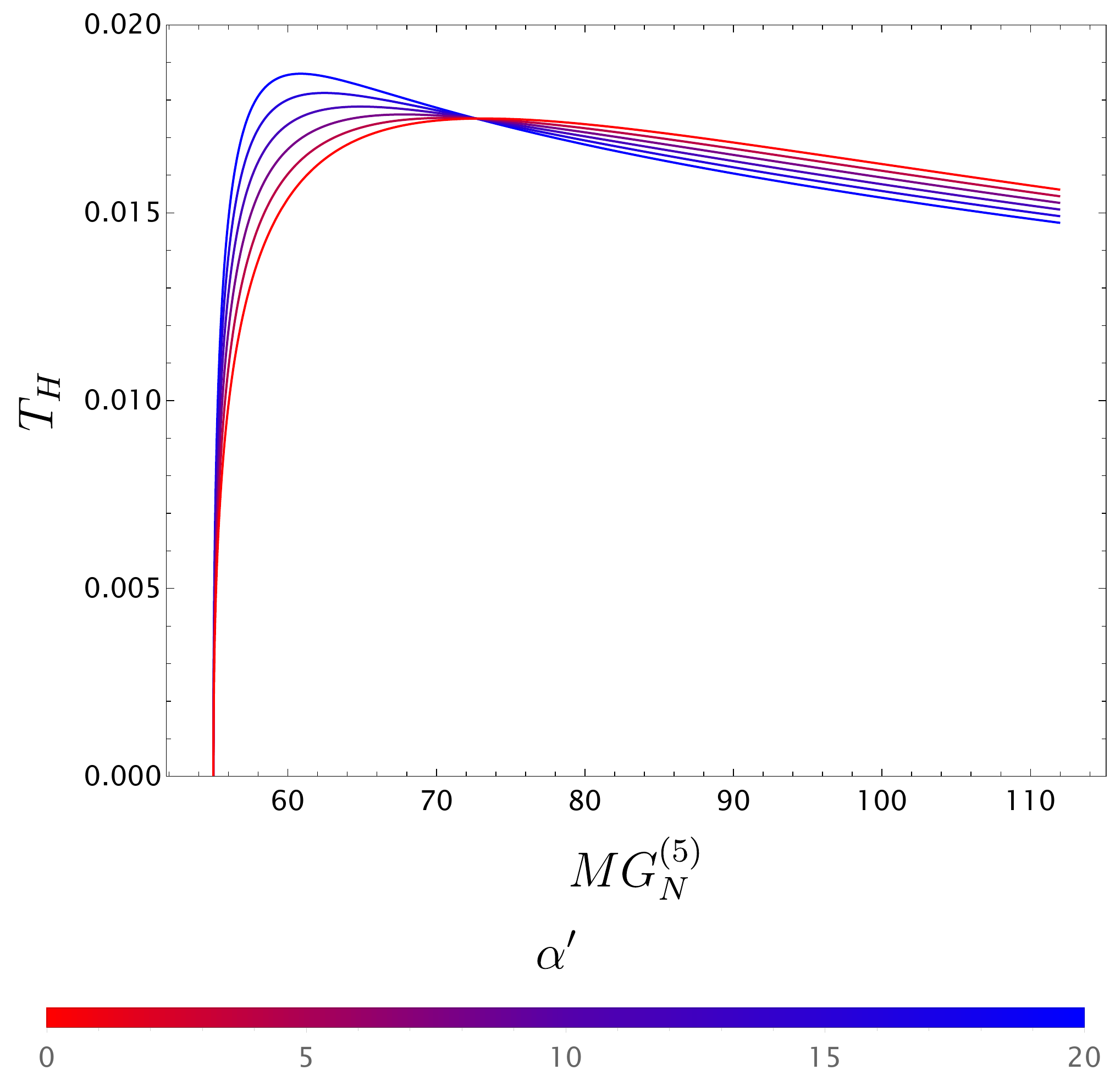}
  \caption{\small The Hawking temperature as a function of the mass for
    $\eta_{+} Q_{+} = 40$, $\eta_- Q_{-} = 20$, $\eta_0 Q_{0} = 10$, for different values of
    $\alpha'$ ($\eta_{(i)}$ are the same of  Eq.~(\ref{eq:Qqbetagamma})). Observe that, for these charges, the black hole becomes
    supersymmetric in the extremal limit.}
\label{fig:P4}
\end{figure}

\begin{figure}[t!]
  \centering \includegraphics[width=0.6\textwidth]{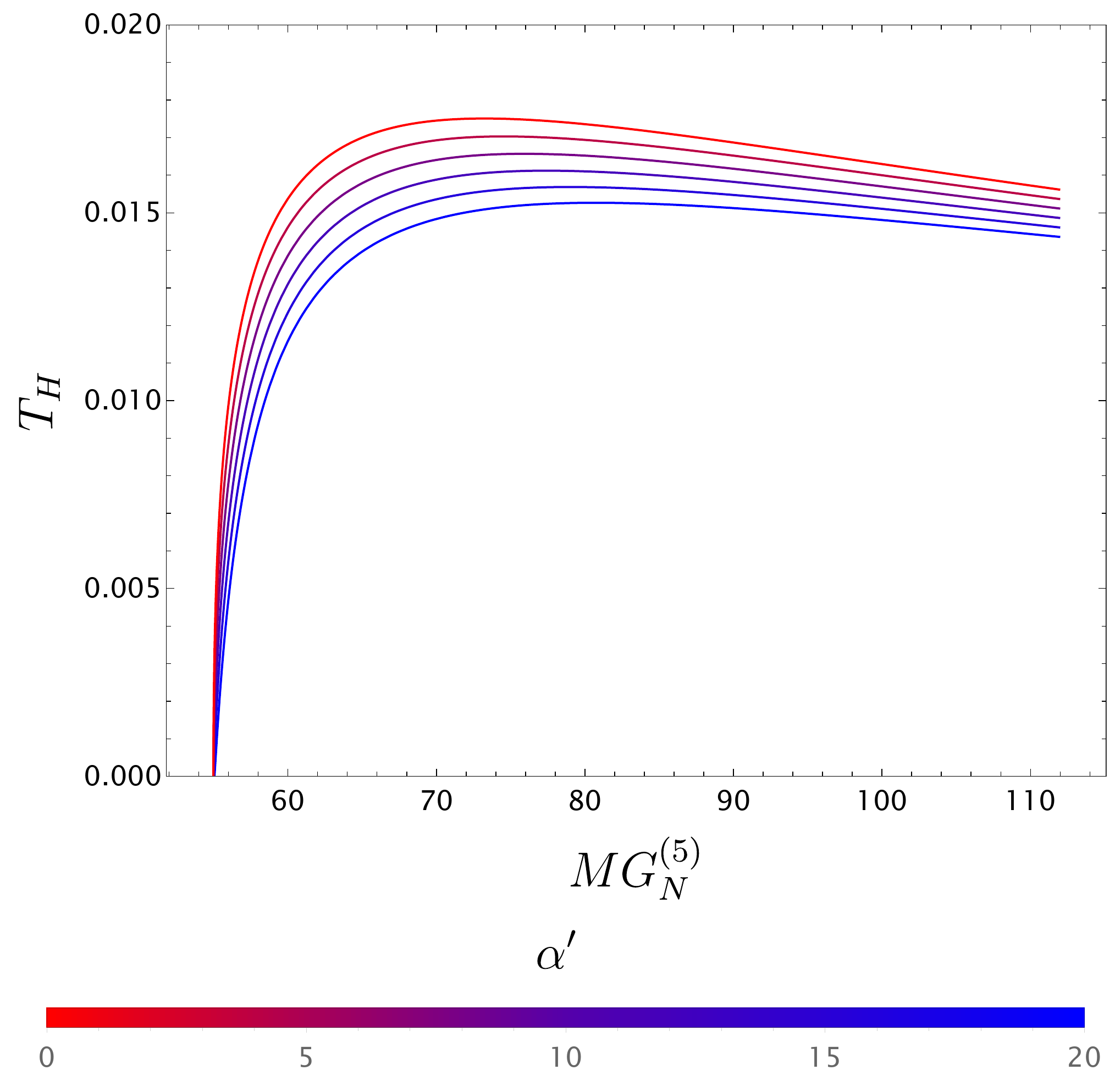}
  \caption{\small The Hawking temperature as a function of the mass for
     $\eta_{+} Q_{+} = 40$, $\eta_- Q_{-} = -20$, $\eta_0 Q_{0} = 10$ for different values of
    $\alpha'$. Observe that, for these charges, the black hole does not become
    supersymmetric in the extremal limit.}
\label{fig:P5}
\end{figure}

\begin{figure}[t!]
  \centering \includegraphics[width=0.6\textwidth]{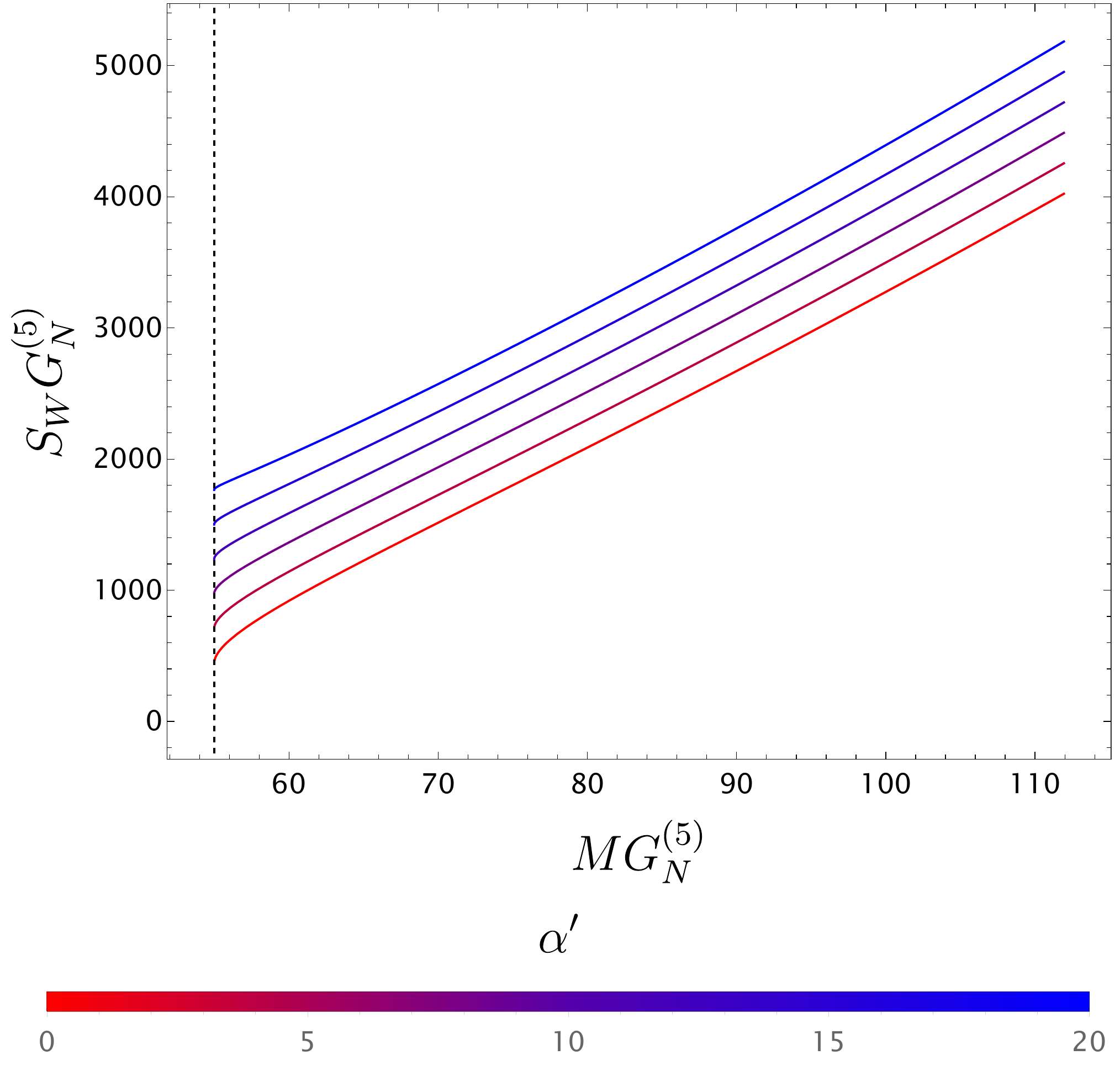}
  \caption{\small The Wald entropy as a function of the mass for
     $\eta_{+} Q_{+} = 40$, $\eta_- Q_{-} = 20$, $\eta_0 Q_{0} = 10$ for different values of
    $\alpha'$. Observe that, for these charges, the black hole becomes
    supersymmetric in the extremal limit.}
\label{fig:P6}
\end{figure}

\begin{figure}[t!]
  \centering \includegraphics[width=0.6\textwidth]{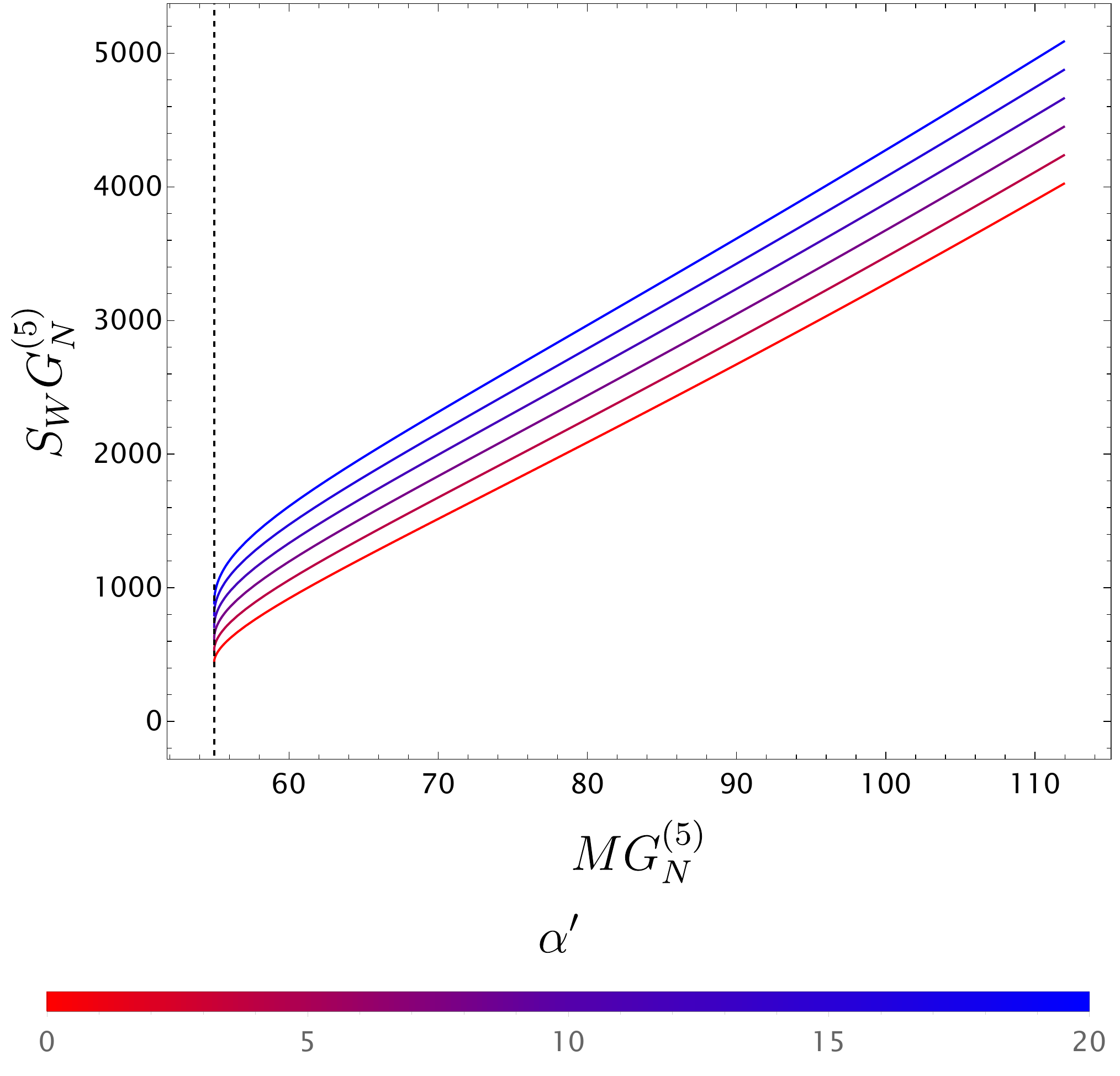}
  \caption{\small The Wald entropy as a function of the mass for
    $\eta_{+} Q_{+} = 40$, $\eta_- Q_{-} = -20$, $\eta_0 Q_{0} = 10$ for different values of
    $\alpha'$. Observe that, for these charges, the black hole does not become
    supersymmetric in the extremal limit.}
\label{fig:P7}
\end{figure}

To end this section, we present Figs.~\ref{fig:P4} and \ref{fig:P5} in which
the Hawking temperature of the $1^{\rm st}$-order solutions is plotted versus
the mass for two typical choices of charges, one of which leads to a
supersymmetric solution in the extremal case and the other to a
non-supersymmetric one, and Figs.~\ref{fig:P6} and \ref{fig:P7} in which the
Wald entropy is plotted versus the mass for several values of $\alpha'$. For
smaller values of the charges and larger values of $\alpha'$ the slope of the
curves changes in the near-extremal region, but in those conditions the
effective action we are using is not valid and it would be necessary to add
higher-order terms.

\subsection{$\omega = 0$}

Setting $\omega = 0$ we recover values of the thermodynamical variables
computed in Ref.~\cite{Cano:2021nzo} for extremal black holes:\footnote{It has
  to be taken into account that the parameters $q_{i}$ in this paper
  correspond to the parameters $\hat{q}_{i}$ in Ref.~\cite{Cano:2021nzo}.}

\begin{subequations}
  \begin{align}
    M^{(1)}_{\rm ext}
    & =
      \frac{\pi}{4 G_{N}^{(5)}}\left(q_{0}+q_{-}+q_{+}\right) \,,
    \\
    & \nonumber \\
    T_{H}^{(1)}{}_{\rm ext}
    & =
      0 \,,
    \\
    & \nonumber \\
    S_{W}^{(1)}{}_{\rm ext}
    & =
      \frac{\pi^{2}}{2G_{N}^{(5)}} \sqrt{q_{+}q_{-}\left[ q_{0}
      +\alpha(2 +s_{+}s_{-}) \right]}\,.
  \end{align}
\end{subequations}

\section{Conclusions}
\label{sec-conclusions}

In this paper we have found the $1^{\rm st}$-order corrections in $\alpha'$ to
the non-extremal Strominger-Vafa black hole of Ref.~\cite{Horowitz:1996ay} in
the framework of the Bergshoeff-de Roo heterotic superstring effective theory
\cite{Bergshoeff:1989de}. In order to determine the integration constants we
have demanded regularity and that the physical charges (mass and electric charges) are not modified by the corrections. This greatly simplifies many expressions.
Furthermore, we have used in a crucial way the invariance of the solution
under $\alpha'$-corrected T~duality transformations
\cite{Bergshoeff:1995cg,Elgood:2020xwu} to simplify the calculations and as a
consistency check of the final results.

We have also used the first law as a consistency check of the solution: the
relation $\partial S/\partial M = 1/T$ involves quantities which are computed
from the metric alone, $M$ and $T$, and a quantity, the Wald entropy $S$,
which depends on the metric as well as on the matter fields through a
complicated formula. Thus, it provides a non-trivial consistency check of the
solution and of the entropy formula itself. The recovery of the entropy
computed in Ref.~\cite{Cano:2021nzo} adds confidence to it. This is the second
instance in which the first law has been explicitly checked for
$\alpha'$-corrected non-extremal heterotic black holes \cite{Cano:2019ycn}.

Another interesting aspect of our results is the emergence of the Regge slope
parameter, $\alpha'$, as a thermodynamical variable. While, on general
grounds, this was to be expected
\cite{Kastor:2010gq,Hajian:2021hje,Ortin:2021ade}, it is not so easy to argue
that there is a $(d-1)$-form potential associated to this parameter that
allows us to understand it as just another charge \cite{Meessen:2022hcg}. This
aspect of our results needs further research to be properly understood.

Almost by definition, the $\alpha'$ corrections do not lead to dramatic
changes in the solution: if they did, they would typically be associated to
large curvatures and, therefore, the calculation and addition of higher order
corrections in $\alpha'$ would be needed to obtain a reliable string solution.
This is why the removal or cloaking of singularities by the corrections, a
widely expected and interesting quantum or stringy effect which in the case
studied in Ref.~\cite{Cano:2018aod} produces a globally regular black hole,
requires the computation of higher-order corrections to be fully confirmed.

The corrections to the thermodynamical quantities, though, are important even
if small since a microscopic model dual to the black hole should be able to
reproduce them. In spite of the efforts devoted to this problem over the
years, such a model has not yet been found. There have been some proposals for
the near-extremal black holes which we intend to study in forthcoming work,
taking into account that, strictly speaking, in that regime the
thermodynamical description of black holes might break down
\cite{Preskill:1991tb}.\footnote{This point has been recently revisited in
  Refs.~\cite{Iliesiu:2020qvm,Heydeman:2020hhw}, which have clarified how the
  naive breakdown of the thermodynamical description at low temperatures
  pointed out in \cite{Preskill:1991tb} is resolved.}

The extension of our results to the 4-dimensional non-extremal stringy black
hole of Ref.~\cite{Horowitz:1996ac}, taking into account the signs of charges
that lead to non-supersymmetric extremal limits studied in
Ref.~\cite{Cano:2021nzo} is another direction of research that we intend to
pursue in the near future. For equal charges, these solutions give an
embedding of the non-extremal Reissner-Nordstr\"om black hole different from
the one studied in Ref.~\cite{Cano:2019ycn} and it will be interesting to see
how different these corrections are for the same metric.

Finally, it would be interesting to consider the effect of rotation. Neutral rotating black holes were recently studied by Refs.~\cite{Cano:2021rey,Agurto-Sepulveda:2022vvf}, but obtaining charged rotating black hole solutions of the heterotic superstring effective action at first order in $\alpha'$ still remains an open problem.

\section*{Acknowledgments}

TO would like to thank Manus Visser for interesting conversations.  This work
has been supported in part by the grants IFT Centro de Excelencia Severo Ochoa
CEX2020-001007-S and by PID2019-110058GB-C22 funded by
MCIN/AEI/10.13039/501100011033 and by ERDF. The work of PAC is supported by a
postdoctoral fellowship from the Research Foundation - Flanders (FWO grant
12ZH121N). The work of AR is supported by a postdoctoral fellowship associated
to the MIUR-PRIN contract 2017CC72MK003. MZ is supported by the fellowship
LCF/BQ/DI20/11780035 from ``La Caixa'' Foundation (ID 100010434). TO wishes to
thank M.M.~Fern\'andez for her permanent support.

\appendix

\section{The heterotic superstring effective action}
\label{app-action}

For the sake of completeness, in this appendix we briefly review the effective
action of the heterotic superstring theory at first order in $\alpha'$ that we
have used in this paper and which is, essentially, that of
Ref.~\cite{Bergshoeff:1989de} adapted to the conventions of
\cite{Ortin:2015hya} and with the non-Abelian gauge fields consistently
truncated.\footnote{The $\alpha'$ corrections to that action we first studied
  in Refs.~\cite{Gross:1986mw,Metsaev:1987zx,Bergshoeff:1989de}. More recent,
  and relevant, studies of those corrections can be found in
  Refs.~\cite{Chemissany:2007he,Baron:2017dvb,Baron:2018lve}.}

In order to describe this action is convenient to start by defining the
$0^{\rm th}$-order 3-form field strength of the Kalb-Ramond 2-form $\hat{B}$

\begin{equation}
  \hat{H}^{(0)}
  =
  d\hat{B}\,,
\end{equation}

$\hat{H}^{(0)}$ is combined with the Levi-Civita spin connection
$\hat{\omega}^{\hat{a}}{}_{\hat{b}}$ into the definition of the torsionful
spin connection

\begin{equation}
  \hat{\Omega}_{(-)}{}^{\hat{a}}{}_{\hat{b}}
  \equiv
  \hat{\omega}^{\hat{a}}{}_{\hat{b}}
  -\frac{1}{2}\hat{H}^{(0)}_{\hat{c}}{}^{\hat{a}}{}_{\hat{b}}\,\hat{e}^{\hat{c}}\,,
\end{equation}

\noindent
whose curvature 2-form and Chern-Simons 3-form are defined by 

\begin{subequations} 
  \begin{align}
  \hat{R}_{(-)}{}^{\hat{a}}{}_{\hat{b}}
  & =
  d\hat{\Omega}_{(-)}{}^{\hat{a}}{}_{\hat{b}}
  -\hat{\Omega}_{(-)}{}^{\hat{a}}{}_{\hat{c}} \wedge
    \hat{\Omega}_{(-)}{}^{\hat{c}}{}_{\hat{b}}\,,
    \\
    & \nonumber\\
  \label{def:chern-simons}
    \hat{\Omega}^{\mathrm{L}}_{(-)}
    & =
    d\hat{\Omega}_{(-)}{}^{\hat{a}}{}_{\hat{b}}\wedge
    \hat{\Omega}_{(-)}{}^{\hat{b}}{}_{\hat{a}}
    -\frac{2}{3} \hat{\Omega}_{(-)}{}^{\hat{a}}{}_{\hat{b}} \wedge
    \hat{\Omega}_{(-)}{}^{\hat{b}}{}_{\hat{c}} \wedge \hat{\Omega}_{(-)}{}^{\hat{c}}{}_{\hat{a}} \,.
 \end{align}
\end{subequations} 

The latter is used into the definition of the first-order 3-form KR field
strength\footnote{Sometimes we will use $H^{(1)}$ to emphasize the fact that
  we are dealing with the first-order field strength.}

\begin{equation}
\label{def:kalb-ramond}
\hat{H}
=
d\hat{B}+ \frac{\alpha'}{4} \hat{\Omega}^{\mathrm{L}}_{(-)}\,.
\end{equation}

With these definition, the effective action of the heterotic string to first
order in $\alpha'$ is given by\footnote{As it is written, this action
  implicitly contains terms of second order in $\alpha'$. They are included
  for the sake of convenience by they should be consistently disregarded in
  all computations.}

\begin{equation}
  \label{eq:action}
  \hat{S}
  =
  \frac{\hat{g}_{s}^{2}}{16\pi G_{\rm N}^{(10)}}
  \int d^{10}x
  \sqrt{|\hat{g}|}\, e^{-2\hat{\phi}}\,
  \left[
    \hat{R} -4(\partial\hat{\phi})^{2} +\frac{1}{2\cdot 3!}\hat{H}^{2}
    -\frac{\alpha'}{8}\hat{R}_{(-)\, \hat{\mu}\hat{\nu}}{}^{\hat{a}}{}_{\hat{b}}
    \hat{R}_{(-)}{}^{\hat{\mu}\hat{\nu}\, \hat{b}}{}_{\hat{a}}
  \right]\,,
\end{equation}

Setting $\alpha'=0$ one gets the $0^{\rm th}$-order action.

\subsection{Equations of motion}
\label{sec-eom}

In order to derive the first-order equations of motion it is convenient to use
the lemma of Ref.~\cite{Bergshoeff:1989de} which implies that the equations of
motion can be obtained by varying the action only with respect to
\textit{explicit} occurrences of the fields (\textit{i.e.}, ignoring them when
they occur inside the torsionful spin connection). The resulting equations of
motion are

\begin{subequations}
\begin{align}
  \label{eq:Einsteinequation}
  \hat{R}_{\hat{\mu}\hat{\nu}}
  -2\hat{\nabla}_{\hat{\mu}}\partial_{\hat{\nu}}\hat{\phi}
  +\frac{1}{4}\hat{H}_{\hat{\mu}\hat \rho\hat \sigma}\hat{H}_{\hat{\nu}}{}^{\hat \rho\hat \sigma}
  -\frac{\alpha'}{4}\hat{R}_{(-)\, \hat{\mu}\hat{\rho}}{}^{\hat{a}}{}_{\hat{b}}
  \hat{R}_{(-)\, \hat{\nu}}{}^{\hat{\rho}\,\hat{b}}{}_{\hat{a}}
  & =
    0\,,
  \\
  &  \nonumber \\
  \label{eq:dilatonequation}
  (\partial \hat{\phi})^{2} -\frac{1}{2}\hat{\nabla}^{2}\hat{\phi}
  -\frac{1}{4\cdot 3!}\hat{H}^{2}
  +\frac{\alpha'}{32}\hat{R}_{(-)\,\hat{\mu}\hat{\nu}}{}^{\hat{a}}{}_{\hat{b}}
  \hat{R}_{(-)}{}^{\hat{\mu}\hat{\nu}\,\hat{b}}{}_{\hat{a}}
  & =
 0\,,
  \\
                   &  \nonumber \\
  \label{eq:Krequation}
  d\left(e^{-2\hat{\phi}}\star\!\hat{H}\right)
  & = 
    0\,.
\end{align}
\end{subequations}

As usual, if one expresses the KR field in terms of its 3-form field strength,
one must also impose the KR Bianchi identity that ensures the local existence
of the KR 2-form potential. It takes the form

\begin{equation}
  \label{eq:bianchi} 
  d\hat{H}
  =
  \frac{\alpha'}{4}\hat{R}_{(-)}{}^{\hat{a}}{}_{\hat{b}}\wedge\hat{R}_{(-)}{}^{\hat{b}}{}_{\hat{a}} \,.
\end{equation}

\section{The non-extremal, $0^{\rm th}$-order solutions}
\label{app-zerothorder}

The ansatz for the $0^{\rm th}$-order 10-dimensional solutions can be
expressed in terms of 4 functions of $\rho$

\begin{equation}
  \label{eq:4functions}
  \mathcal{Z}_{0}\,,\,\,\mathcal{Z}_{+}\,,\,\,\mathcal{Z}_{-}\,,\,\,W\,,
\end{equation}

\noindent
and 5 constants

\begin{equation}
  \beta_{0}\,,\,\,\beta_{+}\,,\,\,\beta_{-}\,,\,\,
  \hat{\phi}_{\infty}\,,\,\,k_{\infty}\,,
\end{equation}

\noindent
plus the integration constants present in the 4 functions.

According to the no-hair conjectures, the independent physical parameters of
the solution are expected to be just 6: the mass $M$, 3 Abelian charges that
we will denote by $q_{0},q_{+},q_{-}$ and the two real moduli
$\hat{\phi}_{\infty},k_{\infty}$. The equations of motion will impose three
relations between these 5 constants and the 4 integration constants associated
to the 4 functions.

In terms of these functions and constants, the ansatz takes the form

\begin{subequations}
\label{eq:3-charge10dsolutionzerothorder}
\begin{align}
d\hat{s}^{2}
  & =
    \frac{1}{\mathcal{Z}_{+}\mathcal{Z}_{-}}Wdt^{2}
    -\mathcal{Z}_{0}(W^{-1}d\rho^{2}+\rho^{2}d\Omega_{(3)}^{2})
    \nonumber \\
  & \nonumber \\
  & 
    -\frac{k_{\infty}^{2}\mathcal{Z}_{+}}{\mathcal{Z}_{-}}
    \left[dz+\beta_{+}k_{\infty}^{-1}
    \left(\mathcal{Z}^{-1}_{+}-1\right)dt\right]^{2}
    -dy^{m}dy^{m}\,,
\hspace{.5cm}
m=1,\ldots,4\,,
\label{eq:d10metriczerothorder}
  \\
  & \nonumber \\
  \hat{H}^{(0)}
  & = 
    \beta_{-}d\left[ k_{\infty}\left(\mathcal{Z}^{-1}_{-}-1\right)
    dt \wedge dz\right]
    +\beta_{0}\rho^{3}\mathcal{Z}'_{0}\omega_{(3)}\,,
  \\
  & \nonumber \\
  e^{-2\hat{\phi}}
  & =
    e^{-2\hat{\phi}_{\infty}}
    \mathcal{Z}_{-}/\mathcal{Z}_{0}\,,
\end{align}
\end{subequations}

\noindent
where a prime indicates derivation with respect to the radial coordinate
$\rho$,

\begin{subequations}
  \begin{align}
  d\Omega^{2}_{(3)}
  & =
  \frac{1}{4}\left[ (d\psi+\cos{\theta}d\varphi)^{2}
    + d\Omega^{2}_{(2)} \right]\,,
    \\
    & \nonumber \\
  d\Omega^{2}_{(2)}
  & =
  d\theta^{2}+\sin^{2}{\theta}d\varphi^{2}\,,
  \end{align}
\end{subequations}

\noindent
are, respectively, the metrics of the round 3- and 2-spheres of unit radii and

\begin{equation}
\omega_{(3)}= \frac{1}{8}d\cos{\theta}\wedge d\varphi\wedge d\psi\,,
\end{equation}

\noindent
is the volume 3-form of the former.

This ansatz reduces to the one for extremal black holes (in particular, for
the Strominger-Vafa black hole \cite{Strominger:1996sh}) when the so-called
``blackening factor'' $W$ is absent or, equivalently, $W=1$.

The equations of motion are solved at $0^{\rm th}$ order in $\alpha'$ for
\cite{Horowitz:1996ay}

\begin{equation}
\label{eq:Zs3-chargezerothorder}
\begin{aligned}
\mathcal{Z}_{0}
  & =
1+\frac{q_{0}}{\rho^{2}}\,, 
\\
& \\
\mathcal{Z}_{-}
  & =
1+\frac{q_{-}}{\rho^{2}}\,, 
\\
&  \\
\mathcal{Z}_{+}
  & =
1+\frac{q_{+}}{\rho^{2}}\,, 
\\
&  \\
W
& =
1+\frac{\omega}{\rho^{2}}\,,
\end{aligned}
\end{equation}

\noindent
where asymptotic flatness and the standard normalization of the metric at
spatial infinity have already been imposed, leaving just 4 integration
constants

\begin{equation}
q_{0}\,,\,\,q_{+}\,,\,\,q_{-}\,,\,\,\omega\,,
\end{equation}

\noindent
which are related to the other 5 constants by the following 3 relations

\begin{equation}
\label{eq:omegaqbetarelation}  
  \beta_{i} = s_{i}\sqrt{1-\frac{\omega}{q_{i}}}\,,
  \hspace{1cm}
  s_{i}^{2}=1\,,
  \hspace{1cm}
  i=0,+,-\,.
\end{equation}

\noindent
This leaves us, as expected, with 4 independent physical parameters (3 charges
and the mass) plus 2 independent moduli: the asymptotic value of the
10-dimensional dilaton $\hat{\phi}_{\infty}$ and the radius of the compact
dimension parametrized by the coordinate $z$ measured in string length units
$k_{\infty}\equiv R_{z}/\ell_{s}$.

When $\omega=0$ the solution describes the extremal, 3-charge black holes
considered in Ref.~\cite{Cano:2021nzo}. In this limit all the $\beta$
parameters square to one, $\beta_{0,\pm}^{2}=1$, and they are just give the
signs of the charges.

When $\omega\neq 0$, the $\beta$ parameters are complicated functions of the
physical parameters which are very complicated to determine explicitly.

The $\omega>0$ case is related to the $\omega<0$ by the coordinate
transformation

\begin{equation}
  \label{eq:omegatominusomegacoordinatetransformation}
\rho^{2} \rightarrow \rho^{2}-\omega\,,
\end{equation}

\noindent
and a redefinition of the parameters

\begin{equation}
  \label{eq:omegatominusomega0}
  q_{i}
  \rightarrow
  q_{i}+\omega\,,
  \hspace{1cm}
  \beta_{i}
  \rightarrow
  \frac{q_{i}}{q_{i}+\omega}\beta_{i}\,
\end{equation}

\noindent
followed by

\begin{equation}
  \label{eq:omegatominusomega}
  \omega\rightarrow -\omega\,,
\end{equation}

\noindent
where now $\omega$ is taken to be negative.

Observe that this transformation preserves the relations
Eqs.~(\ref{eq:omegaqbetarelation}) and also the products
$\beta_{(i)} q_{(i)}$.

Therefore, we may take $\omega<0$ with no loss of generality. Negative values
of the parameters $q_{0\,\pm}$ can also be related to positive ones by similar
transformations. However, these transformations shift $\omega$ by positive
quantities and we may end up violating the assumed negativity of
$\omega$. Thus, we will not make any assumptions concerning the signs of those
parameters although it is well known that they have to be strictly positive if
we want to obtain regular black hole in the extremal limit. Nevertheless, we
will consider both the $\omega>0$ and $\omega<0$ cases because we will use the
results obtained in each case will be used in the main body of the paper.

In order to study the black-hole spacetime described by this solution, we
rewrite it first in 5-dimensional form.

\subsection{The 5-dimensional form of the solutions}
\label{sec-d5-5dimensional}

Upon trivial dimensional reduction on T$^{4}$ (parametrized by the coordinates
$y^{m}$) we get a 6-dimensional solution which is identical to the
10-dimensional one except for the absence of the $-dy^{m}dy^{m}$ term in the
metric. A non-trivial compactification on the S$^{1}$ parametrized by the
coordinate $z$ using the relations in \cite{Elgood:2020xwu} at $0^{\rm th}$ order in
$\alpha'$ gives the 5-dimensional string-frame action and a further rescaling
of the metric

\begin{equation}
  g_{\mu\nu}
  =
  e^{4(\phi-\phi_{\infty})/3}\tilde{g}_{\mu\nu}
  \equiv
  e^{4\tilde{\phi}/3}\tilde{g}_{\mu\nu}\,,
\end{equation}

\noindent
gives the (modified) Einstein-frame action \cite{Maldacena:1996ky}. It is also
convenient to dualize the 2-form $B$ into another 1-form $D$, with field
strength $K=dD$ through the relation\footnote{We use $C$ for the winding
  vector coming from the reduction of the 2-form $\hat{B}$ to distinguish it
  from the 2-form $B$. Its 2-form field strength is $G=dC$.}

\begin{equation}
  \label{eq:HdualK}
  H
  =
  e^{8\tilde{\phi}/3}\star K\,.
\end{equation}

In terms of these variables, but removing the tildes of the Einstein-frame
metric and dilaton for the sake of simplicity, the action and the solutions we
are considering take the form \cite{Elgood:2020mdx}

\begin{equation}
\label{eq:heteroticd5Einsteinframedual}
\begin{aligned}
  S[e^{a},\phi,k,A,C,D]
  & =
  \frac{1}{16\pi G_{N}^{(5)}} \int 
  \left[\star (e^{a}\wedge e^{b}) \wedge R_{ab}
    +\tfrac{4}{3}d\phi\wedge \star d\phi
      +\tfrac{1}{2}k^{-2}dk \wedge \star dk
  \right.
  \\
  & \\
  & \hspace{.5cm}
    -\tfrac{1}{2}k^{2}e^{-4\phi/3}F\wedge \star F
    -\tfrac{1}{2}k^{-2}e^{-4\phi/3}G\wedge \star G
    -\tfrac{1}{2}e^{8\phi/3}K\wedge \star K
  \\
  & \\
  & \hspace{.5cm}
  \left.
        -F\wedge G\wedge C
  \right]\,,
\end{aligned}
\end{equation}

\noindent
and

\begin{subequations}
\label{eq:5dsolution}
\begin{align}
ds^{2}
  & =
f^{2}Wdt^{2}
-f^{-1}(W^{-1}d\rho^{2}+\rho^{2}d\Omega_{(3)}^{2})\,,
\,\,\,\,\,\,\,
f^{-3}= \mathcal{Z}_{+}\mathcal{Z}_{-}\mathcal{Z}_{0}\,,
\\
& \nonumber \\
  F
  & =
    d\left[\beta_{+}k_{\infty}^{-1}
    \left(\mathcal{Z}^{-1}_{+}-1\right)dt\right] \,,
  \\
& \nonumber \\
  G
  & =
    d\left[\beta_{-}k_{\infty}\left(\mathcal{Z}^{-1}_{-}-1\right)dt\right]\,,
  \\
& \nonumber \\
  K
  & =
    d\left[-\beta_{0}\left(\mathcal{Z}^{-1}_{0}-1\right)dt\right]\,,
  \\
& \nonumber \\
e^{-2\phi}
  & =
\sqrt{\mathcal{Z}_{+}\mathcal{Z}_{-}}/\mathcal{Z}_{0}\,,
\\
& \nonumber \\
  k
  & =
   k_{\infty} \sqrt{\mathcal{Z}_{+}/\mathcal{Z}_{-}}\,.
\end{align}
\end{subequations}

The mass of these solutions is given by

\begin{equation}
  \label{eq:Mzerothorder}
  M
  =
  \frac{3\pi}{8G_{N}^{(5)}}
  \left\{-\omega +\tfrac{2}{3}\left(q_{+}+q_{-}+q_{0}\right)\right\}\,.
\end{equation}

\noindent
Observe that the positivity of the mass is compatible with a range of negative
values of the parameters $q_{0\,\pm}$. Furthermore, this formula, which is
valid for the $\omega>0$ and $\omega<0$ cases is invariant under the
transformations Eqs.~(\ref{eq:omegatominusomega0}) and
(\ref{eq:omegatominusomega}).

The physical charges of these solutions can be defined as\footnote{These
  definitions are equivalent to those in Ref.~\cite{Cano:2021nzo} (excluding
  the $\ell_{s}$ factor introduced there for the sake of convenience) because
  the second terms in the integrands do not contribute in these
  configurations.}

\begin{subequations}
  \label{eq:5dchargesdef}
  \begin{align}
    Q_{+}
    & =
      \frac{-1}{16\pi G_{N}^{(5)}}
      \int_{S^{3}_{\infty}}\left\{ e^{-4\phi/3}k^{2}\star F +G\wedge C\right\}
      =
      \frac{\pi}{4G_{N}^{(5)}}k_{\infty}\beta_{+}q_{+}\,,
    \\
    & \nonumber \\
    Q_{-}
    & =
      \frac{-1}{16\pi G_{N}^{(5)}}
      \int_{S^{3}_{\infty}} \left\{ e^{-4\phi/3}k^{-2}\star G +F\wedge C\right\}
      =
      \frac{\pi}{4G_{N}^{(5)}}k^{-1}_{\infty}\beta_{-}q_{-}\,,
    \\
    & \nonumber \\
    Q_{0}
    & =
      \frac{-1}{16\pi G_{N}^{(5)}}
      \int_{S^{3}_{\infty}} \left\{ e^{8\phi/3}\star K +F\wedge B\right\}
      =
      -\frac{\pi}{4G_{N}^{(5)}}\beta_{0}q_{0}\,.
  \end{align}
\end{subequations}

These definitions guarantee that the result does not change when the
integration surface is displaced across a region where the equations of motion
are satisfied and there are no sources. In the language of
Ref.~\cite{Marolf:2000cb}, these are \textit{localized} charges. In
particular, we get the same value when we integrate on the event horizon and
when we integrate at spatial infinity (S$^{3}_{\infty}$). Furthermore, observe
that, once the constants $q_{0\, \pm}$ have been chosen, the signs of the
physical charges are essentially determined by the signs of the $\beta$
parameters. Finally, observe that, as the mass Eq.~(\ref{eq:Mzerothorder}),
the physical charges are invariant under the transformations of the parameters
Eqs.~(\ref{eq:omegatominusomega0}) and (\ref{eq:omegatominusomega}).

In these coordinates, if none of the parameters $q_{0,\pm}$ vanishes, the
horizons are located at the zeroes of $g_{tt}$, \textit{i.e.}~at
$\rho = \rho_{H}= 0$ (outer, event, horizon) for $\omega>0$ and at
$\rho = \rho_{H}=0,\sqrt{-\omega}$ (inner, Cauchy and outer, event, horizons,
respectively) for $\omega<0$. Therefore, the inner horizon is no covered by
the coordinate patch in the $\omega >0$.

If some of the parameters $q_{0,\pm}$ are negative, $g_{tt}$ blows up at
$\rho^{2}=|q|$ where $q$ is any of the negative parameters. The event horizon
covers all these singularities if $|\omega|>|q|$ for all the negative
parameters. This condition does not guarantee the positivity of the mass when
there is more than one negative $q$. In that case, it has to be imposed
independently, though. We will assume that these two properties hold.

We are going to study the thermodynamics in the $\omega >0$ and $\omega <0$
cases separately for the sake of convenience.

\subsubsection{The $\omega >0$ case}

In this case, the radius of the horizon, defined as

\begin{equation}
  \label{eq:radiusofthehorizon}
  R_{H}
  \equiv
  \left.\sqrt{|g_{E\,\theta\theta}|}\right|_{\rho\rightarrow \rho_{H}}\,,  
\end{equation}

\noindent
is given by

\begin{equation}
  \label{eq:RH0omegapositive}
  R_{H}
  =
  \left|\rho/f^{1/2} \right|_{\rho \rightarrow -|\omega|^{1/2}}
  =
 \left( q_{+}q_{-}q_{0}\right)^{1/6}\,,
\end{equation}

\noindent
and the Hawking temperature and Bekenstein-Hawking entropy of the black hole
are given by the expressions

\begin{subequations}
  \begin{align}
    \label{eq:zerothorderHtemperatureomegapositive}
  T_{H}
  & =
  \frac{1}{2\pi} \frac{\omega}{\sqrt{q_{+}q_{-}q_{0}}}\,,
  \\
    & \nonumber \\
    \label{eq:zerothorderBHentropyomegapositive}
  S_{BH}
  & =
  \frac{\pi^{2}}{2G_{N}^{(5)}}\sqrt{q_{+}q_{-}q_{0}}\,,
  \end{align}
\end{subequations}

\noindent
so that

\begin{equation}
  \label{eq:TSproductomegapositive}
  T_{H}S_{BH}
  =
  \frac{\pi \omega}{4G_{N}^{(5)}}\,.
\end{equation}

We can use this relation to rewrite Eq.~(\ref{eq:Mzerothorder}) as follows:

\begin{equation}
  \begin{aligned}
    M
    & =
    \tfrac{3}{2}S_{BH}T_{H}+
  \frac{3\pi}{8G_{N}^{(5)}}
  \left\{-2\omega +\tfrac{2}{3}\left(q_{+}+q_{-}+q_{0}\right)\right\}
  \\
  & \\
    & =
    \tfrac{3}{2}S_{BH}T_{H}
    +\frac{\pi}{4G_{N}^{(5)}}
    \left[(q_{+}-\omega)+(q_{-}-\omega)+(q_{0}-\omega)\right]
  \\
  & \\
    & =
    \tfrac{3}{2}S_{BH}T_{H}
    +\frac{\pi}{4G_{N}^{(5)}}
    \left[(1-\frac{\omega}{q_{+}})q_{+}+(1-\frac{\omega}{q_{-}})q_{-}
      +(1-\frac{\omega}{q_{0}})q_{0}\right]
  \\
  & \\
    & =
    \tfrac{3}{2}S_{BH}T_{H}
    +\frac{\pi}{4G_{N}^{(5)}}
    \left(k_{\infty}^{-1}\beta_{+}Q_{+}+k_{\infty}\beta_{+}Q_{-}-\beta_{0}Q_{0}\right)\,,
  \end{aligned}
\end{equation}

\noindent
where we have used the definitions of the charges Eqs.~(\ref{eq:5dchargesdef})
and the relations Eqs.~(\ref{eq:omegaqbetarelation}).

Comparing this equation with the Smarr formula that can be derived by
homogeneity arguments or via Komar integrals \cite{kn:TOM}

\begin{equation}
  \label{eq:zerothSmarrformula}
  M
  =
  \tfrac{3}{2}S_{BH}T_{H} + \Phi^{+}Q_{+} +\Phi^{-}Q_{-} + \Phi^{0}Q_{0}\,.
\end{equation}

\noindent
we conclude that the electric potentials evaluated on the horizon are given by

\begin{equation}
  \Phi^{+}
  =
  k^{-1}_{\infty}\beta_{+}\,,
  \hspace{1cm}\
  \Phi^{-}
  =
  k_{\infty}\beta_{-}\,,
  \hspace{1cm}
  \Phi^{0}
  =
  -\beta_{0}\,.
\end{equation}

This identification can be checked by a explicit calculation of the form of
the first law, which we perform in Appendix~\ref{sec-checkingfirstlawzeroth}
and also using the definitions 

\begin{equation}
  \label{eq:definitionselectrostaticpotentials}
  \imath_{\partial_{t}}F
  \equiv
  d\Phi^{+}\,,
  \hspace{1cm}
  \imath_{\partial_{t}}G
  \equiv
  d\Phi^{-}\,,
  \hspace{1cm}
    \imath_{\partial_{t}}K
  \equiv
  d\Phi^{0}\,.
\end{equation}

The Smarr formula can also be read as a Bogomol'nyi-type bound

\begin{equation}
  \label{eq:relation}
  M
  -\left( \Phi^{+}Q_{+} +\Phi^{-}Q_{-} + \Phi^{0}Q_{0}\right)
  =
  \tfrac{3}{2}S_{BH}T_{H}
  \geq
  0\,,
\end{equation}

\noindent
which is saturated in the extremal limit $S_{BH}T_{H}\sim \omega=0$. This
suggests the following definition for the central charge

\begin{equation}
  \mathcal{Z}
  \equiv
  \left( \Phi^{+}Q_{+} +\Phi^{-}Q_{-} + \Phi^{0}Q_{0}\right)\,.
\end{equation}

The occurrence in this expression of the $\beta$ parameters, which, in
general, depend on the mass and charges, is not so surprising: in the extremal
limit, they are just signs and the expression is quite standard. In the
non-extremal case, we, actually, expect the bound Eq.~(\ref{eq:relation}) to
take a more complicated form that allows for different extremal limits
associated to the different skew-eigenvalues of the central charge matrix of
at supergravity theory with 16 supercharges.\footnote{See, for instance, the
  discussions in
  Refs.~\cite{Kallosh:1992ii,Bergshoeff:1996gg,Lozano-Tellechea:1999lwm}.}

\subsubsection{The $\omega <0$ case}

In this case, the radius of the event horizon, the Hawking temperature and the
Bekenstein-Hawking entropy of the black hole are given by the expressions

\begin{subequations}
  \begin{align}
  \label{eq:RH0omeganegative}
    R_{H}
    & =
      \left[ (q_{+}-\omega)(q_{-}-\omega)(q_{0}-\omega)\right]^{1/6}\,,
    \\
    &   \nonumber \\
    \label{eq:zerothorderHtemperatureomeganegative}
    T_{H}
    & =
      \frac{1}{2\pi} \frac{-\omega}{\sqrt{ (q_{+}-\omega)(q_{-}-\omega)(q_{0}-\omega)}}\,,
    \\
    & \nonumber \\
    \label{eq:zerothorderBHentropyomeganegative}
    S_{BH}
    & =
      \frac{\pi^{2}}{2G_{N}^{(5)}}\sqrt{ (q_{+}-\omega)(q_{-}-\omega)(q_{0}-\omega)}\,,
  \end{align}
\end{subequations}

As expected, these expressions could have been obtained from those of the
$\omega>0$ case Eqs.~(\ref{eq:RH0omegapositive}),
(\ref{eq:zerothorderHtemperatureomegapositive}) and
(\ref{eq:zerothorderBHentropyomegapositive}) applying the transformations
Eqs.~(\ref{eq:omegatominusomega0}) and (\ref{eq:omegatominusomega}).

The relation Eq.~(\ref{eq:TSproductomegapositive}) is also preserved with the
replacement $\omega \rightarrow -\omega$ and we can use it, as we did in the
$\omega>0$ case, in Eq.~(\ref{eq:Mzerothorder}) to derive the Smarr formula
Eq.~(\ref{eq:zerothSmarrformula}). Now, the potentials take the form

\begin{equation}
  \Phi^{+}
  =
  \frac{k^{-1}_{\infty}}{\beta_{+}}\,,
  \hspace{1cm}\
  \Phi^{-}
  =
  \frac{k_{\infty}}{\beta_{-}}\,,
  \hspace{1cm}
  \Phi^{0}
  =
  -\frac{1}{\beta_{0}}\,,
\end{equation}

\noindent
in agreement with the definitions
Eqs.~(\ref{eq:definitionselectrostaticpotentials}).

It is interesting to compute the same quantities for the inner horizon at
$\rho=0$ as well, assuming none of the $q$s are negative. The result is

\begin{subequations}
  \begin{align}
    R_{H\, inner}
    & =
      \left( q_{+}q_{-}q_{0}\right)^{1/6}\,,
    \\
    &   \nonumber \\
    T_{H\,inner}
    & =
      \frac{1}{2\pi} \frac{\omega}{\sqrt{q_{+}q_{-}q_{0}}}\,,
    \\
    & \nonumber \\
    S_{BH\, inner}
    & =
      \frac{\pi^{2}}{2G_{N}^{(5)}}\sqrt{q_{+}q_{-}q_{0}}\,,
  \end{align}
\end{subequations}

\noindent
and we find

\begin{equation}
  \label{eq:TSproductszerothorder}
-T_{H\, inner}S_{BH\,inner} =T_{H}S_{BH} = -\frac{\pi \omega}{4G_{N}^{(5)}}\,.  
\end{equation}

Observe that the charges are equal on both horizons, as follows from the
definition we have used (the charges are \textit{localized}).

\subsection{The Reissner-Nordstr\"om-Tangherlini solution}
\label{app-RNT}

An important particular case of the solutions we are considering is the one in
which the 3 charge parameters are equal $q_{0}=q_{+}=q_{-}\equiv q$. In this
case, the three 5-dimensional vector fields are proportional, the scalar
fields are constant everywhere and the metric becomes that of the
5-dimensional Reissner-Nordstr\"om-Tangherlini black hole
\cite{Tangherlini:1963bw}. 

In this case it is not difficult to express the integration constants
$q,\omega,\beta=\beta_{0}$ in terms of the physical parameters $M,Q=Q_{0}$
($Q_{\pm}=k_{\infty}^{\pm 1}Q_{0}$):

\begin{subequations}
  \label{eq:integrationparametersversusphysicalparameters}
  \begin{align}
    \omega
    & =
      \pm \frac{8 G_{N}^{(5)}}{3\pi}\sqrt{M^{2}-9Q^{2}}\,,
    \\
    & \nonumber \\
    q
    & =
      \frac{12 G_{N}^{(5)}}{\pi} \frac{Q^{2}}{M\mp \sqrt{M^{2}-9Q^{2}}}\,,
    \\
    & \nonumber \\
    \beta
    & =
      \frac{3Q}{M\pm \sqrt{M^{2}-9Q^{2}}}\,,
      \hspace{1cm}
      \beta_{\pm}= s_{\pm} \beta\,.
  \end{align}
\end{subequations}

\noindent
Observe that

\begin{equation}
\beta(\omega > 0) = 1/\beta(\omega <0)\,.  
\end{equation}

\noindent
This property is necessary for the electrostatic potential $\Phi$, which we
have identified with $\beta$ in the $\omega > 0$ case and with $1/\beta$ in
the $\omega <0$ case to be a physical quantity whose expression does not
depend on the coordinates chosen.

Using these relations we can express the radius of the horizon, the Hawking
temperature, the Bekenstein-Hawking entropy and the electrostatic potential in
terms of the physical parameters

\begin{subequations}
  \begin{align}
    \label{eq:RHRNT}
    R_{H}
    & =
      \left(\frac{12 G_{N}^{(5)}}{\pi}\right)^{1/2}
      \frac{|Q|}{\left(M- \sqrt{M^{2}-9Q^{2}}\right)^{1/2}}\,,
    \\
    & \nonumber \\
        \label{eq:zerothorderHtemperatureRNT}
    T_{H}
    & =
      \left(\frac{3}{4\pi G_{N}^{(5)}}\right)^{1/2}
      \frac{\sqrt{M^{2}-9Q^{2}}}{\left(M+ \sqrt{M^{2}-9Q^{2}}\right)^{3/2}}\,,
    \\
    & \nonumber \\
        \label{eq:zerothorderHentropyRNT}
    S_{BH}
    & =
      \tfrac{2}{3} \left(\frac{4\pi G_{N}^{(5)}}{3}\right)^{1/2}
      \left(M+ \sqrt{M^{2}-9Q^{2}}\right)^{3/2}\,,
    \\
    & \nonumber \\
    \Phi
    & =
      \frac{3Q}{M+ \sqrt{M^{2}-9Q^{2}}}\,.      
  \end{align}
\end{subequations}

The extremal limit of this solution is

\begin{equation}
M= 3|Q|\,.  
\end{equation}

\subsection{T~duality}
\label{sec-Tdualityzeroth}

The dimensionally reduced $0^{\rm th}$-order action
Eq.~(\ref{eq:heteroticd5Einsteinframedual}) is invariant under the
transformations

\begin{equation}
  \label{eq:Tdualityzerothorder}
  A \leftrightarrow C\,,
  \hspace{1cm}
  k \leftrightarrow 1/k\,.
\end{equation}

\noindent
which interchange the KK vector $A$ with the winding vector $C$ and invert the
KK scalar that measures the radius of the compact dimension. This symmetry is
manifestation at the level of the effective-field theory action of the
T~duality of the heterotic superstring compactified on a circle, which
interchanges KK and winding modes and inverts the compactification
radius. When the above transformations are reexpressed in terms of the
higher-dimensional variables (components of the metric, KR field and dilaton)
\cite{Bergshoeff:1994dg} they take the form of the famous ``Buscher T~duality
rules'' \cite{Buscher:1987sk,Buscher:1987qj}.

It is not difficult to see that the effect of the above transformations on the
$0^{\rm th}$-order solutions Eqs.~(\ref{eq:3-charge10dsolutionzerothorder}) is
equivalent to the following transformation of some of the parameters of the
solutions only:

\begin{equation}
  \label{eq:zerothorderTdualityparameters}
  \beta_{+}  \leftrightarrow \beta_{-}\,,
  \hspace{1cm}
  q_{+}  \leftrightarrow q_{-}\,,
  \hspace{1cm}
  k_{\infty} \leftrightarrow 1/k_{\infty}\,.
\end{equation}

Thus, Eqs.~(\ref{eq:3-charge10dsolutionzerothorder}), understood as a family
of solutions with arbitrary values of these parameters, is invariant under
T~duality (or self-dual) and it cannot be extended any further using it.
Since the Einstein metric is invariant under T~duality, all the geometric
properties of the solution (temperature, entropy, first law) must also be
invariant under T~duality and it can be readily seen that, indeed, the
temperature and entropy are invariant under the above transformations.

\section{Checking the first law at zeroth order in $\alpha'$}
\label{sec-checkingfirstlawzeroth}

We are going to assume $\omega<0$, for simplicity.  In order to check the first
law at $0^{\rm th}$ order in $\alpha'$ we can use the following relations between
the parameters of the solution $\omega,q_{0,\pm},\beta_{0,\pm}$ and the
physical parameters $M,Q_{0,\pm}$

\begin{subequations}
  \begin{align}
    \label{eq:wqbeta}
    \omega
    & =
      q_{0,\pm}(1-\beta_{0,\pm}^{2})\,,
    \\
    & \nonumber \\
    \label{eq:Mwq}
    M
    & =
      \frac{3\pi}{8G_{N}^{(5)}}
      \left\{-\omega +\tfrac{2}{3}\left(q_{+}+q_{-}+q_{0}\right)\right\}\,.
    \\
    & \nonumber \\
    \label{eq:Qqbetagamma}
    \eta_{(i)}Q_{(i)}
    & =
      \beta_{(i)}q_{(i)}\,,
  \end{align}
\end{subequations}

\noindent
where $\eta_{0,\pm}$ are parameters that depend on $G_{N}^{(5)}$ and
$k_{\infty}$ and where we have chosen $\omega<0$.

The basic idea is to find how the entropy varies with the physical parameters
using the expression\footnote{It is assumed that the values of the $q_{i}$s
  are such that the entropy is real.}

\begin{equation}
  S
  =
  \frac{\pi^{2}}{2G_{N}^{(5)}}\sqrt{(q_{+}-\omega)(q_{-}-\omega)(q_{0}-\omega)}\,,  
\end{equation}

\noindent
and the above relations between the physical parameters and those in terms of
which the metric and the entropy are given. The coefficient of $\delta M$ must
be $1/T$ with

\begin{equation}
  T
  =
  \frac{1}{2\pi} \frac{-\omega}{\sqrt{(q_{+}-\omega)(q_{-}-\omega)(q_{0}-\omega)}}\,.  
\end{equation}

\noindent
The coefficient of $\delta Q_{i}$ must be the $\Phi^{i}$s

\begin{equation}
  \Phi^{+}
  =
  \frac{k^{-1}_{\infty}}{\beta_{+}}\,,
  \hspace{1cm}\
  \Phi^{-}
  =
  \frac{k_{\infty}}{\beta_{-}}\,,
  \hspace{1cm}
  \Phi^{0}
  =
  -\frac{1}{\beta_{0}}\,.
\end{equation}

Varying the above entropy formula, we get

\begin{equation}
  \label{eq:deltaS1}
  \begin{aligned}
     \delta S
 & =
  \frac{\pi^{2}}{4G_{N}^{(5)}}
  \frac{1}{\sqrt{(q_{+}-\omega)(q_{-}-\omega)(q_{0}-\omega)}}
  \left\{
 \right.
 \\
 & \\
 & \hspace{.5cm}
   -\left[(q_{-}-\omega)(q_{0}-\omega) 
  +(q_{+}-\omega)(q_{0}-\omega) 
  +(q_{+}-\omega)(q_{-}-\omega) \right]
   \delta \omega
 \\
 & \\
 & \hspace{.5cm}
 \left.
 +(q_{-}-\omega)(q_{0}-\omega) \delta q_{+}
  +(q_{+}-\omega)(q_{0}-\omega) \delta q_{-}
  +(q_{+}-\omega)(q_{-}-\omega) \delta q_{0}
    \right\}\,,  
  \end{aligned}
\end{equation}

From Eq.~(\ref{eq:Mwq}) we obtain

\begin{equation}
  \label{eq:deltaomega1}
  \delta \omega
  =
  -\frac{8G_{N}^{(5)}}{3\pi}\delta M  +\tfrac{2}{3}\sum \delta q_{i}\,.
\end{equation}

Using the relations Eqs.~(\ref{eq:Qqbetagamma}) in Eqs.~(\ref{eq:wqbeta}) we
get

\begin{equation}
  \label{eq:deltaqi1}
  \delta q_{i}
  =
  \frac{1}{1+\beta_{i}^{2}} \delta \omega
  +\frac{2\eta_{(i)}\beta_{(i)}}{1+\beta_{i}^{2}} \delta Q_{(i)}\,,
\end{equation}

\noindent
and substituting this result into Eq.~(\ref{eq:deltaomega1}) we get

\begin{equation}
  \label{eq:deltaomega2}
  \delta \omega
  =
  -\frac{8G_{N}^{(5)}}{\pi}X\delta M
  +4X\sum\frac{\eta_{j}\beta_{j}}{1+\beta_{j}^{2}} \delta Q_{j}\,,
\end{equation}

\noindent
where we have defined

\begin{equation}
  X
  \equiv
  \tfrac{1}{3}
  \left(1-\tfrac{2}{3}\sum \frac{1}{1+\beta_{i}^{2}}\right)^{-1}
  = -\frac{1}{\omega} \left(\sum \frac{1}{2q_{i}-\omega}\right)^{-1}\,.
\end{equation}

\noindent
Substituting this result back into Eqs.~(\ref{eq:deltaqi1}) we find

\begin{equation}
  \label{eq:deltaqi2}
  \delta q_{i}
  =
  -\frac{1}{1+\beta_{i}^{2}}\frac{8G_{N}^{(5)}}{\pi}X \delta M
+\frac{1}{1+\beta_{i}^{2}}\left[
  4X  \sum\frac{\eta_{j}\beta_{j}}{1+\beta_{j}^{2}} \delta Q_{j}
  +2\eta_{(i)}\beta_{(i)}\delta Q_{(i)}\right]\,.
\end{equation}

Now, we can substitute Eqs.~(\ref{eq:deltaomega2}) and (\ref{eq:deltaqi2})
into Eq.~(\ref{eq:deltaS1}) and identify the coefficients of $\delta M$ and
$\delta Q_{i}$. The first of these coefficients is given by

\begin{equation}
  \label{eq:1/T}
  \begin{aligned}
1/T
 & =
  2\pi
  \frac{X}{\sqrt{(q_{+}-\omega)(q_{-}-\omega)(q_{0}-\omega)}}
  \left\{
 \right.
 \\
 & \\
 & \hspace{.5cm}
   \left[(q_{-}-\omega)(q_{0}-\omega) 
  +(q_{+}-\omega)(q_{0}-\omega) 
  +(q_{+}-\omega)(q_{-}-\omega) \right]
 \\
 & \\
 & \hspace{.5cm}
 \left.
 -\left[(q_{-}-\omega)(q_{0}-\omega) \frac{1}{1+\beta_{+}^{2}}
  +(q_{+}-\omega)(q_{0}-\omega) \frac{1}{1+\beta_{-}^{2}}
  +(q_{+}-\omega)(q_{-}-\omega) \frac{1}{1+\beta_{0}^{2}}\right]
\right\}
\\
& \\
& =
2\pi
  \frac{X}{\sqrt{(q_{+}-\omega)(q_{-}-\omega)(q_{0}-\omega)}}
  \left\{
 (q_{-}-\omega)(q_{0}-\omega)\frac{\beta_{+}^{2}}{1+\beta_{+}^{2}}
 \right.
 \\
 & \\
 & \hspace{.5cm}
 \left.
  +(q_{+}-\omega)(q_{0}-\omega)\frac{\beta_{-}^{2}}{1+\beta_{-}^{2}}
  +(q_{+}-\omega)(q_{-}-\omega) \frac{\beta_{-}^{2}}{1+\beta_{-}^{2}}
\right\}\,.
  \end{aligned}
\end{equation}

\noindent
Since

\begin{equation}
  \frac{\beta_{i}^{2}}{1+\beta_{i}^{2}}
  =
  \frac{q_{i}-\omega}{2q_{i}-\omega}\,,
\end{equation}

\begin{equation}
  \label{eq:1/T2}
  \begin{aligned}
1/T
& =
2\pi
  \frac{X}{\sqrt{(q_{+}-\omega)(q_{-}-\omega)(q_{0}-\omega)}}
  \left\{
 (q_{+}-\omega)(q_{-}-\omega)(q_{0}-\omega)\sum\frac{1}{2q_{i}-\omega}
\right\}
\\
& \\
& =
  2\pi
  X\sqrt{(q_{+}-\omega)(q_{-}-\omega)(q_{0}-\omega)}
  \sum\frac{1}{2q_{i}-\omega}
\\
& \\
& =
  -\frac{2\pi}{\omega}
  \sqrt{(q_{+}-\omega)(q_{-}-\omega)(q_{0}-\omega)}\,,
  \end{aligned}
\end{equation}

\noindent
in agreement with the first law.

Now, let us compute the coefficient of $\delta Q_{+}$, for instance, which
should be equal to $-\Phi^{+}/T$, according to the first law

\begin{equation}
  \begin{aligned}
    -\Phi^{+}/T
    & =
  \frac{\pi^{2}}{4G_{N}^{(5)}}
  \frac{2\eta_{+}\beta_{+}}{(1+\beta_{+}^{2})\sqrt{(q_{+}-\omega)(q_{-}-\omega)(q_{0}-\omega)}}
  \left\{
 \right.
 \\
 & \\
 & \hspace{.5cm}
   -\left[(q_{-}-\omega)(q_{0}-\omega) 
  +(q_{+}-\omega)(q_{0}-\omega) 
  +(q_{+}-\omega)(q_{-}-\omega) \right]
   2X
 \\
 & \\
 & \hspace{.5cm}
 +\left[\frac{(q_{-}-\omega)(q_{0}-\omega)}{1+\beta_{+}^{2}} 
  +\frac{(q_{+}-\omega)(q_{0}-\omega)}{1+\beta_{-}^{2}}  
  +\frac{(q_{+}-\omega)(q_{-}-\omega)}{1+\beta_{0}^{2}} 
\right]2X
 \\
 & \\
 & \hspace{.5cm}
 \left.
+(q_{-}-\omega)(q_{0}-\omega)
\right\}
\\
& \\
    & =
  \frac{\pi^{2}}{4G_{N}^{(5)}}
  \frac{2\eta_{+}\beta_{+}}{(1+\beta_{+}^{2})\sqrt{(q_{+}-\omega)(q_{-}-\omega)(q_{0}-\omega)}}
  \left\{
 \right.
 \\
 & \\
 & \hspace{.5cm}
 -2X\left[\frac{(q_{-}-\omega)(q_{0}-\omega)\beta_{+}^{2}}{1+\beta_{+}^{2}} 
  +\frac{(q_{+}-\omega)(q_{0}-\omega)\beta_{-}^{2}}{1+\beta_{-}^{2}}  
  +\frac{(q_{+}-\omega)(q_{-}-\omega)\beta_{0}^{2}}{1+\beta_{0}^{2}} 
\right]
 \\
 & \\
 & \hspace{.5cm}
 \left.
+\frac{(q_{+}-\omega)(q_{-}-\omega)(q_{0}-\omega)}{(q_{+}-\omega)}
\right\}
\\
& \\
    & =
  \frac{\pi^{2}}{4G_{N}^{(5)}}
  \frac{2\eta_{+}\beta_{+}}{(1+\beta_{+}^{2})\sqrt{(q_{+}-\omega)(q_{-}-\omega)(q_{0}-\omega)}}
  \left\{
 \right.
 \\
 & \\
 & \hspace{.5cm}
 -2X(q_{+}-\omega)(q_{-}-\omega)(q_{0}-\omega)\sum\frac{1}{2q_{i}-\omega}
 \left.
+\frac{(q_{+}-\omega)(q_{-}-\omega)(q_{0}-\omega)}{(q_{+}-\omega)}
\right\}
\\
& \\
    & =
  \frac{\pi^{2}}{2G_{N}^{(5)}}
  \frac{\eta_{+}\beta_{+} \sqrt{(q_{+}-\omega)(q_{-}-\omega)(q_{0}-\omega)}}{(1+\beta_{+}^{2})}
\frac{2q_{i}-\omega}{\omega(q_{+}-\omega)}
\\
& \\
    & =
  \frac{k_{\infty}^{-1}}{\beta_{+}}
  2\pi \frac{\sqrt{(q_{+}-\omega)(q_{-}-\omega)(q_{0}-\omega)}}{\omega}\,,
  \end{aligned}
\end{equation}

\noindent
again in agreement with the first law with the identification of the electric
potential $\Phi^{+}$ made before.

Since all the expressions are symmetric in the three parameters $q_{i}$ and in
the three charges $Q_{i}$ it is not necessary to check the other two terms in
the first law.


\end{document}